\documentclass[apj,numberedappendix]{emulateapj}

\usepackage[sort&compress]{natbib}
\bibpunct{(}{)}{;}{a}{}{,}
\usepackage{color}
\definecolor{darkgreen}{RGB}{0,142,128}

\usepackage{hyperref}
\hypersetup{colorlinks,citecolor=blue,linkcolor=blue}

\usepackage{amsmath}
\usepackage{amsthm}
\usepackage{amsfonts}
\usepackage{url}
\usepackage{subfigure}
\usepackage{multirow}

\newcommand{\bnab}{\boldsymbol{\nabla}}

\newcommand{\Div}{\bnab\cdot}

\newcommand{\reff}[1]{{#1}}

\begin{document}

\title{Magnetic games between a planet and its host star: the key role
of topology}
\shorttitle{Magnetic games between a planet and its host star: the key role
of topology}
%\title{Three dimensional models of star-planet interactions:\\
%  the influence of magnetic topology}
%\shorttitle{Three dimensional models of star-planet interactions:
%  the influence of magnetic topology}

\author{A. Strugarek}
\affil{D\'epartement de physique, Universit\'e de Montr\'eal, C.P. 6128 Succ. Centre-Ville, Montr\'eal, QC H3C-3J7, Canada}
\affil{Laboratoire AIM Paris-Saclay, CEA/Irfu Universit\'e Paris-Diderot CNRS/INSU, F-91191 Gif-sur-Yvette.}
\email{strugarek@astro.umontreal.ca}
\author{A. S. Brun}
\affil{Laboratoire AIM Paris-Saclay, CEA/Irfu Universit\'e Paris-Diderot CNRS/INSU, F-91191 Gif-sur-Yvette.}
\author{S. P. Matt}
\affil{Astrophysics group, School of Physics, University of Exeter, Stocker Road, Exeter EX4 4QL, UK }
\author{V. R\'eville}
\affil{Laboratoire AIM Paris-Saclay, CEA/Irfu Universit\'e Paris-Diderot CNRS/INSU, F-91191 Gif-sur-Yvette.}

\shortauthors{Strugarek, et al.}
\begin{abstract}
Magnetic interactions between a star and a close-in planet are 
postulated to be a source of enhanced emissions and to play a role in
the secular evolution of the orbital system. Close-in planets generally
orbit in the sub-alfv\'enic region of the stellar wind, which
leads to efficient transfers of energy and angular momentum between
the star and the planet. We model the magnetic
interactions occurring in close-in star-planet systems with
three-dimensional, global, compressible magneto-hydrodynamic
numerical simulations of a planet orbiting in a self-consistent
stellar wind. We focus on the cases of magnetized planets and explore
three representative magnetic configurations. % The magnetic topology is
% found to crucially modify how 
% We focus on the cases of magnetized planets and characterize the
% influence of the magnetic topology on the development of magnetic
% interactions.
% The most and least efficient configurations are
% identified, and we confirm that the Poynting fluxes induced by the
% magnetic interaction can reach values higher than $10^{19}$
% W.
The Poynting
flux originating from the magnetic interactions is an energy source
for enhanced emissions in star-planet systems. Our results
suggest a simple geometrical explanation for ubiquitous on/off
enhanced emissions associated with close-in planets, and confirm that
the Poynting fluxes can reach \reff{powers of the order of $10^{19}$ W}.
Close-in planets are also showed to migrate due to magnetic torques
for sufficiently strong stellar wind magnetic fields. The topology of
the interaction significantly modifies the shape of the magnetic
obstacle that leads to magnetic torques. As a consequence,
the torques can vary by at least an order of magnitude as the magnetic
topology of the interaction varies. % We
% demonstrate that magnetic torques generally have to be considered to
% estimate the migration path of close-in planets. 
\end{abstract}


\maketitle

\section{Introduction}
\label{sec:introduction}

The exoplanets detection techniques favor so far the discovery of
giant, close-in planets that can significantly perturb radial velocity
signals, or lead to deep, well defined transits. Furthermore, $34\%$ of
the known exoplanets\footnote{From the database
  \url{exoplanet.eu}, 01/09/2015.} are in orbit closer than $20 \, R_\star$. Such
planets are expected to strongly interact with their hosts \citep[see,
\textit{e.g.},][]{Cuntz:2000ef}, in a
potentially observable way. The interactions originate from tides,
magnetism, and radiative processes. The proximity of close-in
exoplanets amplifies these effects, which can theoretically lead to
energy and angular momentum exchanges between the star and the planet,
and may have observable signatures.

As a matter of fact, several intriguing observations are associated with close-in
exoplanets. \citet{Shkolnik:2008gw} (and references therein) reported
chromospheric emissions for five
different star-planet couples that correlate with the planetary
orbital period. These correlated emissions were observed to be subject
to an on/off mechanism, possibly originating from the variability of
the stellar magnetic field over time-scales of years, or over the orbital phase of the
planet. The particular case of HD 189733 was recently revisited by
\citet{Pillitteri:2015dy}, who interpreted the excess emission to
result from an infall of planetary material towards the star. The surprising
lack of X-ray emissions of WASP-18 is also thought to result from some
star-planet interactions \citep{2014A&A...567A.128P}, which
is yet to be understood. Nevertheless, it is today clear that the
enhancement (or lack of thereof) of
\reff{chromospheric or coronal X-ray emissions} due to a close-in
planet is situational and highly variable: it does not statistically
induce an observational trend \citep[see][and references
therein]{Miller:2015ih}. Radio and UV emissions from star-planet magnetic
interactions are also intensively researched today
\citep{Griessmeier:2007dm,Fares:2010hq,LecavelierdesEtangs:2013fu,2013MNRAS.428..678T},
as any detection may provide constraints on the hypothetical
planetary magnetic fields \citep[e.g.][]{Zarka:2007fo,Vidotto:2015hw}. 

The statistical distribution of exoplanets also reveals interesting
features. First noted by \citet{Pont:2009ip}, it appears that
hosts of close-in planets tend to rotate more rapidly than non- close-in
planets hosting twins. \reff{This was recently confirmed by
\citet{2015A&A...577A..90M}, although the authors question the original explanation of
\citet{Pont:2009ip} based on tidal interactions because they do not
seem to find a correlation between the anomalous gyrochronological age and the
strength of the expected tidal forces.} Furthermore,
\citet{2013ApJ...775L..11M,2014MNRAS.443.1451L} showed a clear dearth
of close-in exoplanets around fast-rotators with Kepler. Both effects could
be explained by exchanges of angular momentum between stars
and close-in planets, although their detailed mechanism is still debated today.

The aforementioned observations are generally interpreted in terms of
tidal, radiative, or magnetic star-planet interactions. Tides are known to lead to
spin-orbit synchronisation in star-planet systems \citep[for a
review, see][]{Mathis:2013cd}. The angular momentum transport
resulting from star-planet tides can also lead to
planet migration \citep[see
\textit{e.g.}][]{Bolmont:2012go,Zhang:2014iz,2015A&A...574A..39D}
as well as spin-up the host star for close-in planets
\citep{Barker:2011jn,Poppenhaeger:2014be,2015ApJ...807...78F}. Planet-disk
tidal interactions provide as well various migration mechanisms
\citep[for a review, see][]{2014prpl.conf..667B} in the
early stages of stellar systems. The
efficiency of tidal interactions strongly
depends on the internal structure of both the star and the planet, and
its modelling is still today a subject of intense research
\citep{AuclairDesrotour:2014io,2014A&A...566L...9G}. 

Close-in planet are as well subject to
intense radiation from their host that can lead to planetary outflows
\citep[see,
\textit{e.g.},][]{2008SSRv..139..437Y,2014MNRAS.444.3761O,Trammell:2014aa}
and may allow in some cases some planetary material to impact the
stellar chromosphere. In the case of close-in non-magnetized planets,
intense EUV radiation was shown to favor the penetration of the
stellar wind and may be a source of enhanced atmospheric escape
\citep{2015ApJ...806...41C}. \citet{Matsakos:2015aa} classified the different
types of planetary outflows, though, further
investigation is still required to elucidate how such
flows could explain enhanced emissions or statistical trends in the
exoplanet population \citep[see][for a possible link between
hot spots and radiation-induced planetary
outflows]{Pillitteri:2015dy}.  

Magnetic interactions provide another promising mechanism for the transfer of
energy and angular moment between a star and a planet. Close-in planet
generally orbit inside the sub-alfv\'enic region
of the stellar wind, leading to particularly efficient transfers
\citep[see, \textit{e.g.},][]{Cohen:2010jm,Strugarek:2014ab}. In a
pioneering work \citet{Ip:2004ba} modelled such interaction
as a plausible source for additional and localized coronal heating of
close-in planet hosting stars. In this scenario, contrary to
radiative-induced planetary outflows, the energy is
carried away from the planet by alfv\'enic perturbations propagating in
the stellar wind down to the stellar chromosphere. The energetic
transfers occurring due to magnetic interactions in a
star-planet system can be modelled with the concept of
Alfv\'en wings \citep{Neubauer:1998aa}, inspired by similar
planet-satellite magnetic interactions occurring in the solar system
\citep{Goldreich:1969kf,Neubauer:1980in}. Though, in 
star-planet systems, the detailed structure of the wind determines how the
Alfv\'en wings develop. The Poynting flux in 
Alfv\'en wings was quantified by \citet{Saur:2013aa} for the known
exoplanets at that time, using simple 1D stellar wind models. It could
provide a source for the intermittent enhanced
emissions sometimes observed in close-in exo-systems. The magnetic torques
originating from star-planet magnetic interactions were proposed as
well to be a source of planet migration
\citep{Laine:2008dx,Lovelace:2008bl,Vidotto:2010iv,Laine:2011jt,Strugarek:2014ab}
and stellar spin-up
\citep{Cohen:2010jm,Lanza:2010bo,Strugarek:2014ab}. On the contrary, for fast rotating
stars those magnetic torques can spin-down the star, albeit not
efficiently enough to solve the so-called angular momentum problem for
young stars \citep{Bouvier:2015kq}. As in the case of tidal
interactions, it is important to note that 
star-planet magnetic interactions (SPMI) generally depend on
the planet internal composition, and in particular whether or not a dynamo process is able
to sustain an intrinsic magnetic field in its interior
\citep{Strugarek:2014ab}. 
% The energy and
% angular momentum transfers are nonetheless very dependent upon the
% magnetic topology of the interaction
% \citep{Saur:2013aa,Strugarek:2014ab}, 

% In a recent work
% \citep{Strugarek:2014ab}, we characterized the magnetized angular momentum
% transfers in close-in star-planet systems for these various
% interactions cases based on a simplified, 2.5D
% axisymmetric setup. We showed that magnetic torques are indeed able
% to make a close-in planet on secular time-scales. 
% Star-planet magnetic
% interactions occur nevertheless intrinsically in three dimensions, we
% report in this paper numerical simulation results of 3D star-planet
% interactions. 

%The magnetic torques (and associated migration time-scales)
%in these new 3D models are confirmed to be sufficiently strong to induce planet
%migration on secular times-scales. 

This paper focuses on the effects of magnetic topology in the development of
magnetic interactions between a star and a close-in, magnetized
planet. We investigate whether magnetic interactions
can be strong enough to explain enhanced emissions or a statistical
dearth of close-in planet around fast-rotating stars. We study
how Alfv\'en wings develop in self-consistent, global
three-dimensional numerical models of stellar winds in which an orbiting,
magnetized planet is added. We
explore three extreme magnetic topologies of aligned, anti-aligned, and
perpendicular configurations. We systematically characterize the
energy and angular momentum exchanges that occur in each
case and demonstrate the crucial influence of the magnetic topology. 
In Section \ref{sec:model-stell-wind} we describe the modelling approach
we chose for the stellar wind and the magnetized planet. A detailed
study of the Alfv\'en wings that self-consistently develop in our
numerical model is given in Section \ref{sec:alfven-wings}. The magnetic
torques leading to planet migration are characterized in Section
\ref{sec:magnetic-torques} and conclusions are given in Section \ref{sec:conclusions}. 

\section{Stellar wind and planet models}
\label{sec:model-stell-wind}

We use the PLUTO code \citep{Mignone:2007iw} to model star-planet
magnetic interactions. We detail here the system of equations we
solve, our modelling choices for the stellar wind and the planet, and
the numerical methods we use.

\subsection{Magneto-hydrodynamic equations}
\label{sec:magn-hydr-equat}

The PLUTO code solves the following set of compressible, ideal magneto-hydrodynamic (MHD) equations:
\begin{eqnarray}
  \label{eq:mass_consrv_pluto}
  &&\partial_t \rho + \boldsymbol{\nabla}\cdot(\rho \mathbf{v}) = 0 \, \\
  \label{eq:mom_consrv_pluto}
  &&\rho\partial_t\mathbf{v} +
  \rho\mathbf{v}\cdot\boldsymbol{\nabla}\mathbf{v}+\boldsymbol{\nabla} P
  +\mathbf{B}\times\boldsymbol{\nabla}\times\mathbf{B}/(4\pi)
  = \rho\mathbf{a} \, ,
  \\
  \label{eq:ener_consrv_pluto}
  &&\partial_t P +\mathbf{v}\cdot\boldsymbol{\nabla} P + \rho
  c_s^2\boldsymbol{\nabla}\cdot\mathbf{v} = 0 \, ,\\
  \label{eq:induction_pluto}
  &&\partial_t \mathbf{B} - \boldsymbol{\nabla}\times\left(\mathbf{v}\times\mathbf{B}\right)
  = 0 \, , \\
  &&\boldsymbol{\nabla}\cdot{\bf B} = 0\, ,
\end{eqnarray}
where $\rho$ is the plasma density, $\mathbf{v}$ its velocity, $P$ the gas
pressure, $\mathbf{B}$ the magnetic field, $\mathbf{a}$ is composed of the gravity, Coriolis, and centrifugal
forces \reff{(the MHD equations are written
  in a rotating reference frame that is specified in Section \ref{sec:planet-models})}, and $c_s=\sqrt{\gamma\,P/\rho}$ the sound
speed ($\gamma$ is the adiabatic exponent, taken to be the equal
  to the ratio of specific heats). We use an
ideal gas equation of state
\begin{equation}
  \label{eq:EOS}
  \rho\varepsilon = P/\left(\gamma-1\right)\, ,
\end{equation}
where $\varepsilon$ is the internal energy per \reff{unit} mass.

\subsection{Stellar wind models}
\label{sec:stellar-winds}

The base of our MHD modelling approach for stellar wind was originally developed by \citet{Washimi:1993vm}, and
extended later on by
\citet{Keppens:1999tw,Matt:2004kd,Matt:2008bj,Matt:2012ib,Strugarek:2014ab,2015ApJ...798..116R}. 
In this work we further develop this approach by considering three dimensional
stellar winds. We briefly explain here our modelling methodology and refer the
interested reader to the aforementioned works for further details.

We model stellar winds using the MHD approximation to describe the
stellar corona plasma. In order to simplify the model, we do not
describe the heating mechanism of the corona itself
\citep[see][]{Suzuki:2006hi}, and instead consider 
our stellar boundary condition to represent the base of the
corona. We
prescribe there a thermal pressure gradient that drives an outward
accelerating flow, the stellar wind. To mimic the additional complex
heating that occurs in the lower corona and participates to
the physical acceleration of stellar winds,
we choose an effective adiabatic exponent
$\gamma$ close to isothermal (in this work $\gamma = 1.05$). This choice
is made to ensure that our model, when applied to the Sun, achieves
velocities compatible with the slow component of the solar wind
observations at 1~AU \citep{Washimi:1993vm,Matt:2004kd}. 

\begin{deluxetable}{lcc}
  \tablecaption{Stellar winds properties\label{tab:tabw}}
  \tablecolumns{3}
  \tabletypesize{\scriptsize}
  %\tablecomments{The numbers are given for a solar-like mass loss rate.}
  \tablehead{
    \colhead{} &
    \colhead{Dipolar wind} & 
    \colhead{Quadrupolar wind}
}
  \startdata 
  $B_\star$ (equator) [G] & 12.4 & 11. \\
  $\dot{M}$ [10$^{-14}$ $M_\odot$/yr] & 2.04 & 2.11 \\
  $\dot{J}$ [10$^{-11}$ $M_\odot R_\odot^2$/yr$^2$] & 56.0 & 2.55 \\
  $\langle R_A \rangle$ [$R_\star$]  & 18.0 & 3.78
  \enddata
\end{deluxetable}

The stellar wind is then determined by the interplay between the
thermally accelerated flow, the large-scale magnetic structures, the
rotation of the star, and the coronal density \citep[see,
e.g.][]{2015ApJ...798..116R,Reville:2015wq}. This latter parameter
can be conveniently used as an adimensionalization parameter for the
MHD equations. The other parameters are described in terms of
characteristic velocities normalized to the escape velocity $v_{\rm
  esc}=\sqrt{2GM_\star/R_\star}$. The thermal driving of the wind is
considered to be spherically symmetric and set by the sound speed
$c_S=\sqrt{\gamma\,P/\rho}$. The rotation of the star is supposed to
be solid and specified by the rotation speed $v_{\rm rot}=\Omega_\star
R_\star$. Finally, the stellar magnetic field is prescribed by a given
idealized topology (dipole or quadrupole) and an equatorial
Alfv\'en speed $v_A = B_\star/\sqrt{4\pi\rho_\star}$, where $B_\star$
is the stellar magnetic field on the equatorial plane. We consider in
this work that the magnetic pole is aligned with the rotation axis.

We initialize our simulation domain with a spherically symmetric Parker wind
\citep{Parker:1958dn} to which we add a magnetic field of a given
topology and a given normalized Alfv\'en speed $v_A/v_{\rm esc}$. The star is modelled
as an internal spherical boundary condition centered at the middle of
the computation domain. The boundary condition consists of three spherical
layers under the stellar 'surface' in which the Parker wind pressure
gradient, the rotation rate of the star, and its magnetic field are
imposed \citep[see][for full details on these
boundary conditions]{Strugarek:2014fr,Strugarek:2014ab}.

We study two different stellar winds using dipolar and quadrupolar
magnetic fields to explore the effects of the magnetic topology on 
SPMIs. Both winds are driven by a normalized sound speed $c_s/v_{\rm
  esc}=0.222$ corresponding to a coronal temperature of $10^6$~K for a
solar-like star. They both slowly rotate with $v_{\rm rot}/v_{\rm esc}=3.03\,
10^{-3}$. The dipolar case is
characterized by a normalized Alfv\'en speed of $v_A/v_{\rm esc}=1$, and in the
quadrupolar case $v_A/v_{\rm esc}=3$. This ensures that the total pressure at the
planet orbit (dominated by the magnetic pressure, see 
Section \ref{sec:planet-models}) is equivalent in both winds.

Because the coronal base density $\rho_\star$ is used to adimensionalize
the MHD equations, each stellar wind simulation can represent the wind
of different stars. We give in Table \ref{tab:tabw} the magnetic field
strength at the base of our models (along with the wind mass and angular momentum
loss rates, and the generalized Alfv\'en radius $\left\langle R_A
\right\rangle = \sqrt{\dot{J}/\Omega_\star\dot{M}}$) for a coronal
base density chosen
to achieve a solar-like mass loss rate ($\rho_\star=3.2\, 10^{-15}$
g/cm$^3$ in the dipolar case, and $2.8\, 10^{-16}$ g/cm$^3$ in the
quadrupolar case). In the remainder of this paper, the dimensional
quantities are given using these base coronal densities unless stated otherwise.

\subsection{Planet models}
\label{sec:planet-models}

The planet is modelled as a second internal boundary condition inside
the computational domain. We consider here planets in circular orbits
with an orbital plane perpendicular to the stellar rotation axis.
The MHD equations are solved in a 
frame rotating at the keplerian orbital rate of the planet $\Omega_K =
\sqrt{GM_\star/R_{\rm orb}^3}$ (in the limit of a small planetary
mass), where $R_{\rm orb}$ is the orbital radius. The boundary cells composing the
star and the planet are fixed in the rotating frame, simplifying
significantly the numerical setup. 

\begin{figure}[htb]
  \centering
  \includegraphics[width=\linewidth]{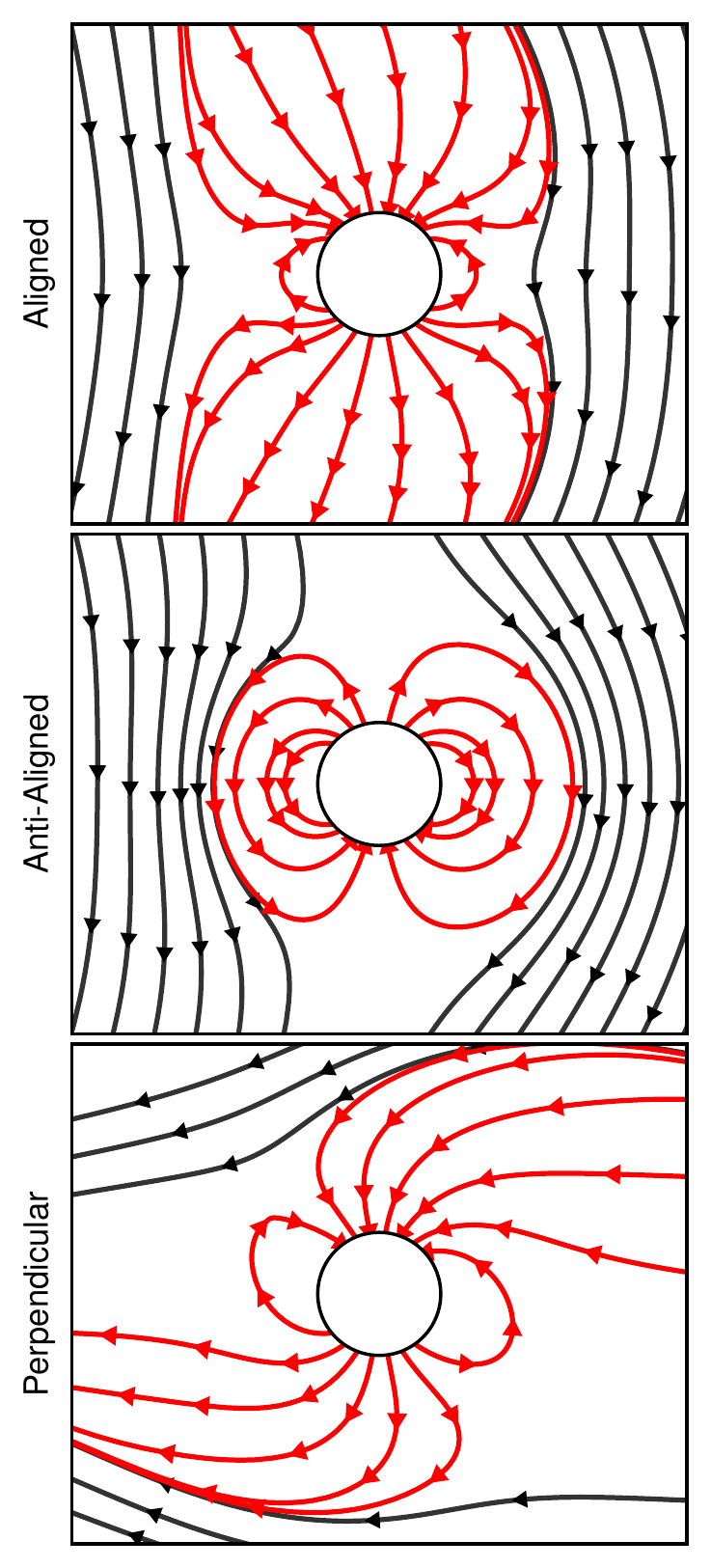}
  \caption{Close-in views of the three magnetic configurations shown
    in this work. The configurations are labelled by the
    orientation of the planetary field (in red) with respect to the local
    stellar wind magnetic field (in black), \textit{i.e.} aligned, anti-aligned,
    and perpendicular, from top to bottom.}
  \label{fig:config_2d}
\end{figure}

In this study we set the planetary mass and radius to $M_P=0.01\,
M_\star$ and $R_P = 0.1\,R_\star$ as in \citet{Strugarek:2014ab}. We
also hold fix the orbital radius $R_{\rm orb} = 5\, R_\star$. We consider 
planets with an intrinsic dipolar magnetic field oriented along
the rotation axis of the host star and simulate three topological
situations. Using the dipolar wind solution, we consider the cases of 
planet with a dipolar field aligned with the \textit{local} magnetic field
(hereafter the 'aligned' case), and of a planet with
an anti-aligned dipolar field (hereafter 'anti-aligned')\footnote{Note
that in \citet{Strugarek:2014ab}, the 'aligned' and 'anti-aligned'
denominations had the opposite meaning: they referred to the
relative orientation of the planetary and stellar dipoles. In this work we
prefer not to use this convention since, for more complex topologies, a
denomination based on the local orientation of the magnetic fields in
the vicinity of the planet is more intuitive.}. With the
quadrupolar wind solution we simulate a planet with a dipolar field parallel to the
stellar rotation axis, and hence perpendicular to the \textit{local} magnetic
field (hereafter the 'perpendicular' case). We show the three
magnetic configurations in Figure \ref{fig:config_2d}, where the stellar wind
field lines are shown in black and the planetary field lines in red. In the 
dipolar wind cases the planetary dipole strength corresponds to an
equatorial field at the planetary surface $B_P=0.9$ G (roughly 5
times less than the jovian magnetic field), and in the quadrupolar wind
cases $B_P=0.25$ G. Such planetary fields ensure extended
planetary magnetospheres, while not constraining too much the
numerical time steps.

The planet is initialized at the beginning of the simulations along
with the stellar wind (see Section \ref{sec:stellar-winds}). We choose
to neglect any kind of atmospheric escape
to focus our study on the magnetic interactions solely \citep[see][for
an overview of the impact of atmospheric escape and planetary 'winds'
in the context of SPMIs]{Matsakos:2015aa}. The density of the planetary
boundary is chosen to be ten times the local wind density, and the
pressure to be $90\%$ of the local thermal pressure of the
wind. This ensures that the planet is a cold, dense obstacle in
the stellar wind from which no 'wind' is triggered. Setting a higher
density and/or a lower pressure changes
very marginally the structure of the planetary magnetosphere close to
the planet boundary, and does not affect our results regarding the
interaction between the planetary magnetosphere, the stellar wind and
its host star. The velocity inside the planet and its magnetosphere is
initially set to 0 in the rotating frame, effectively modelling the
orbital motion of the planet and its magnetosphere. We hence consider
here only planet that are ``tidally-locked'', which is a reasonable
assumption for such close-in planets. \reff{The ``tidal-locking''
  timescale for close-in Jupiter-like planets is estimated to be of
  the order of 0.1-1 Myr \citep[see, \textit{e.g.},][]{2010A&A...516A..64L},
  which is shorter than the typical migration timescales associated
  with magnetic torques we derive in Section
  \ref{sec:magnetic-torques}. We hence consider
in this work that the planet is already tidally-locked on a circular
orbit to focus on the
magnetic torques themselves.} The velocity,
density, and pressure are 
held fixed during the simulation inside the planet boundary and left free to evolve in the
magnetosphere. The magnetic field is maintained to the initial planetary dipole
in the inner $70\%$ of the planet and left free to evolve
elsewhere. The planet boundary condition hence possesses an outer
layer in which only the magnetic field is allowed to change, crudely mimicking
a thick ionosphere. After its initialization, the planetary
magnetosphere dynamically changes until a steady-state is reached, in
which the pressure balance determines the shape of the
magnetosphere. In the case we consider here, the total pressure in the planet vicinity is
dominated by the magnetic pressure, which
ensures that the magnetic interaction develops in the sub-alfv\'enic regime.

\subsection{Numerics}
\label{sec:numerics}

We use the modular PLUTO code \citep{Mignone:2007iw} to solve the
ideal MHD equations (\ref{eq:mass_consrv_pluto}-\ref{eq:EOS}).
The equations
are solved with a second-order, linear spatial interpolation 
coupled to the standard \textit{HLL} Riemann solver
and a \textit{minmod} flux limiter. The variables are updated in time
with a second-order Runge-Kutta method. The solenoidality of the
magnetic field is ensured to machine precision with a constrained
transport method \citep{Evans:1998aa}, in which the face-centered
electromotive forces are arithmetically averaged. 

We solve the MHD equations in cartesian geometry with two internal
boundary conditions inside the domain, modelling the star (Section
\ref{sec:stellar-winds}) and the planet (Section
\ref{sec:planet-models}). \reff{We recall that the equations are
  solved in a rotating frame with the rotation rate $\Omega_K$,
  ensuring that the location of the planet can be held fixed in the
  simulation grid.}
At the domain external boundaries we impose simple
outflow conditions (zero-gradient on all quantities). Because we model
stellar winds, the flow is supersonic and super-alfv\'enic  when it
reaches the \reff{outer} boundaries, hence they have little to no impact on the
general solution which is driven by the internal stellar boundary condition.

The simulation are run on a 490 $\times$ 355 $\times$ 355 cartesian
grid. The cube of size $3 \,R_\star$ enclosing the central star is discretized over
$97$ uniform cells in each direction, and the cube of size $R_\star$ enclosing
the planet over $161$ uniform cells. The remainder
of the simulation domain is filled with stretched grids in the three
directions towards the domain's limits $[-20\,R_\star, 20\,R_\star]^3$.

We stress that we use an ideal set of MHD equations (Section
\ref{sec:magn-hydr-equat}) which implies
that the only dissipative processes occurring in our simulations are
controlled by the numerical scheme 
and the resolution we choose. This limitation is a reasonable trade-off between
numerical simplicity and physical accuracy of our
models. Indeed, in all the cases we consider in this work the
star-planet system quickly reaches a steady-state because the planet is
orbiting in a purely axisymmetric wind (see Section
\ref{sec:stellar-winds}). As a result, the detailed
reconnection process between, \textit{e.g.}, the stellar wind and the
planetary magnetosphere, only marginally influences the final
steady-state. This is confirmed with additional simulations we ran
with half-resolution in the cube of size $R_\star$ around the
planet, which encloses the main reconnection sites. In these
simulations the energy and angular momentum transfers (see Sections
\ref{sec:alfven-wings} and \ref{sec:magnetic-torques}) are decreased by
less than $15\%$. Hence, higher resolution, more accurate simulations would
lead to slightly stronger magnetic interaction, but not qualitatively change
the results presented in this work, as expected. Nevertheless, because
we chose an ideal MHD approach, our model is not suited to
tackle the dynamical response of the interaction to perturbations or
non-axisymmetric structures in the stellar wind. Hence, we only
consider the case of axisymmetric stellar winds in this work. Thanks to PLUTO
modular capabilities, such dynamical processes could be studied more accurately by
taking into account explicit ohmic, Hall, and eventually ambipolar
dissipation \citep[these effects are not included in the public
version of PLUTO yet, for first implementations
see][]{Lesur:2014cc,Nakhaei:2014cg}. These aspects are
beyond the scope of the present work and will be explored
in future studies.

\section{Alfv\'en wings}
\label{sec:alfven-wings}

The concept of Alfv\'en wings goes back to the pioneering work of
\citet{1965JGR....70.3131D}, in the context of satellites moving perpendicularly to a
uniform magnetic field. Such satellite excites alfv\'enic perturbations
that propagate along the magnetic field lines, effectively developing
currents resembling planes' swept-back ``wings''. We detail in this
Section how Alfv\'en wings develop in close-in star-planet systems,
and how the magnetic topology affects their shape and characteristics.

\subsection{Structure of the Alfv\'en wings}
\label{sec:alfv-wings-struct}

Alfv\'en wings develop in close-in star-planet systems due to the
differential motion between the orbiting planet and the rotating,
stellar wind. The orbiting planet
excites magneto-hydrodynamic perturbations in the stellar wind that
propagate along the Alfv\'en characteristics \citep{1965JGR....70.3131D,Neubauer:1998aa,Saur:2013aa}

\begin{equation}
  \label{eq:alfven_characteristics}
  \mathbf{c}_{A}^{\pm} \equiv \mathbf{v} \pm \mathbf{v}_{A}\, .
\end{equation}

These perturbations are a vector of electromagnetic energy and angular
momentum transport between the planet and its environment, the latter being
either the interplanetary medium or the host star. The superposition
of the travelling 
perturbations form what is referred to as an ``Alfv\'en wing'' \citep{Neubauer:1998aa}.
If the local Alfv\'en speed is sufficiently high and the planet is
located inside the Alfv\'en surface of the stellar wind, some of
the perturbations can be reflected at the stellar surface and
reach back the orbiting planet. This extreme case is often
referred to as the \textit{unipolar inductor} case, while the case
where no perturbations reach back the planet is called the \textit{pure
  Alfv\'en wing} case. The star-planet system systematically develops two
Alfv\'en wings, along $c_{A}^{-}$ and $c_{A}^{+}$, located in the
$({\bf v}_0,{\bf B}_w)$ plane, where ${\bf B}_w$ is the stellar wind
magnetic field and ${\bf v}_0={\bf v}_w-{\bf v}_k$ is the differential
motion between the planet and the wind
(${\bf v}_k=R_{\rm orb}\Omega_K {\bf e}_\varphi$ is the keplerian velocity). 

\begin{figure*}[htb]
  \centering
  \includegraphics[width=0.4\linewidth]{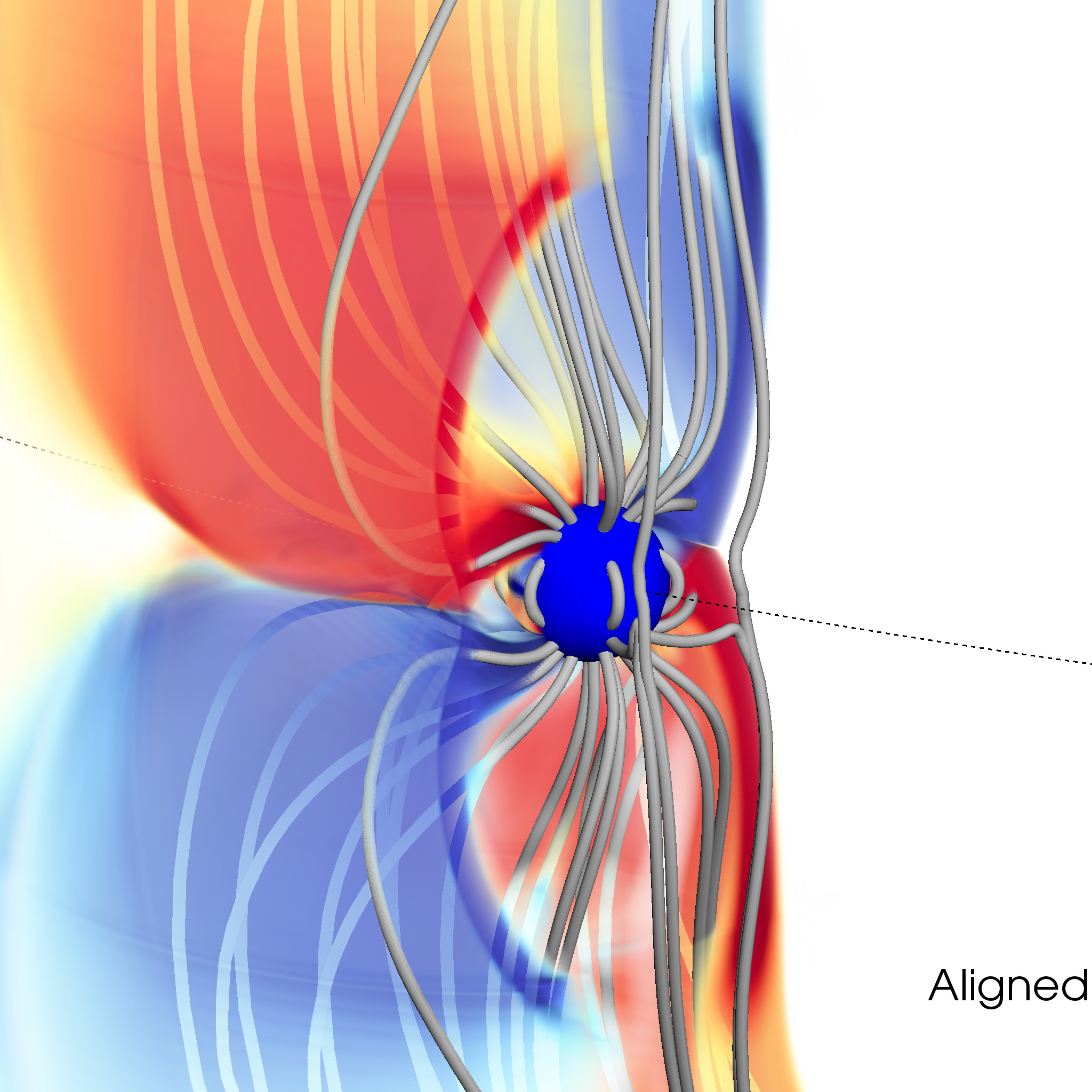}
  \includegraphics[width=0.4\linewidth]{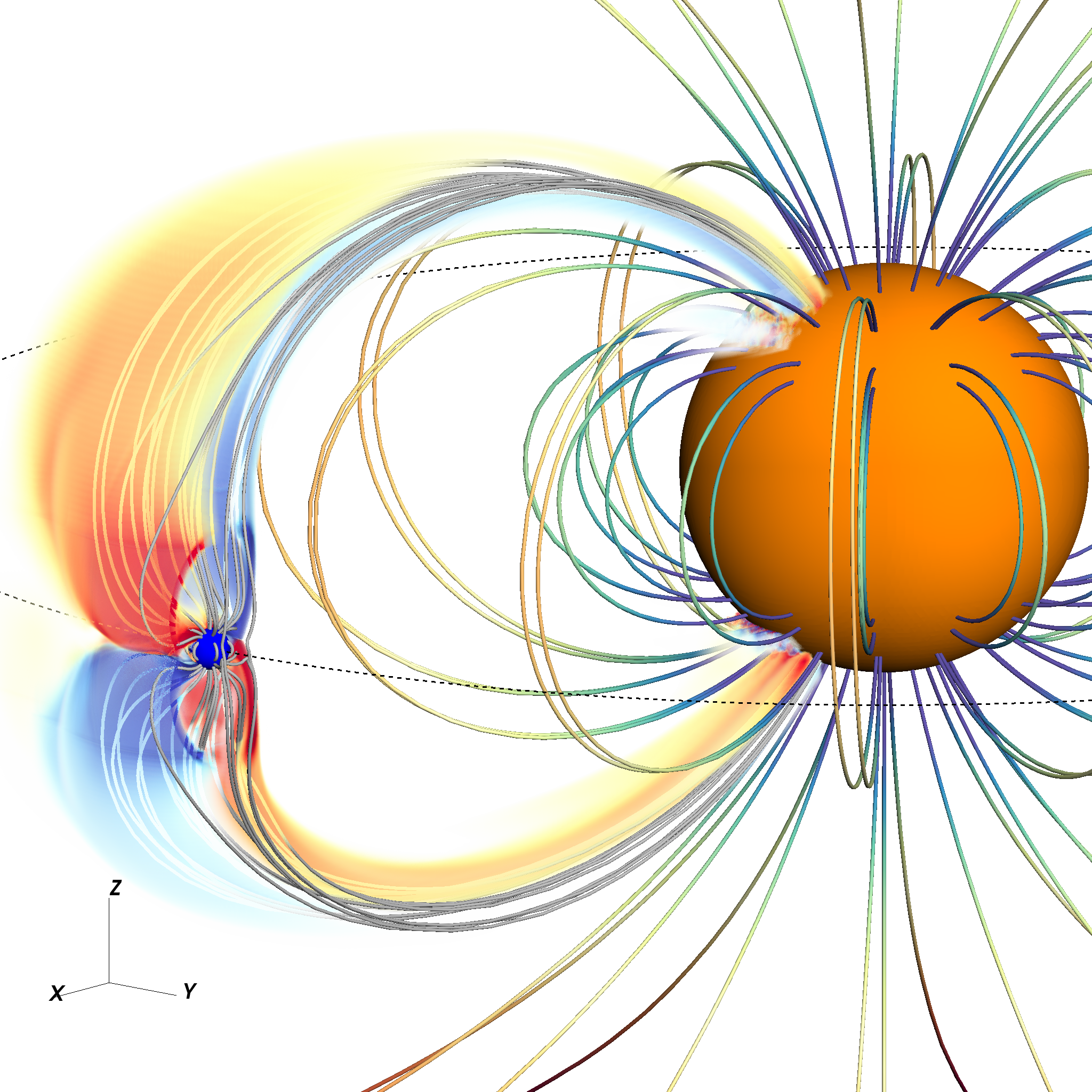}
  \includegraphics[width=0.4\linewidth]{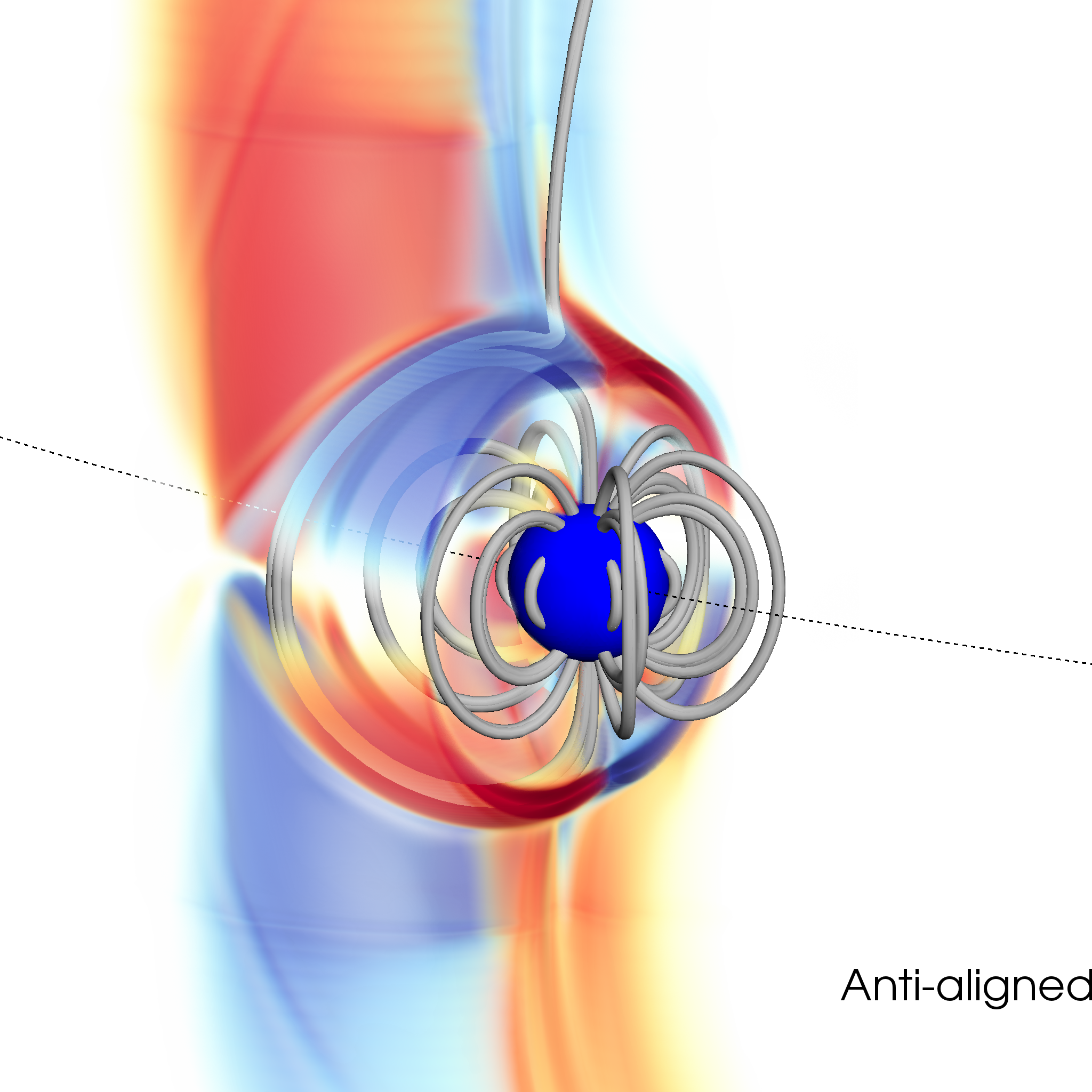}
  \includegraphics[width=0.4\linewidth]{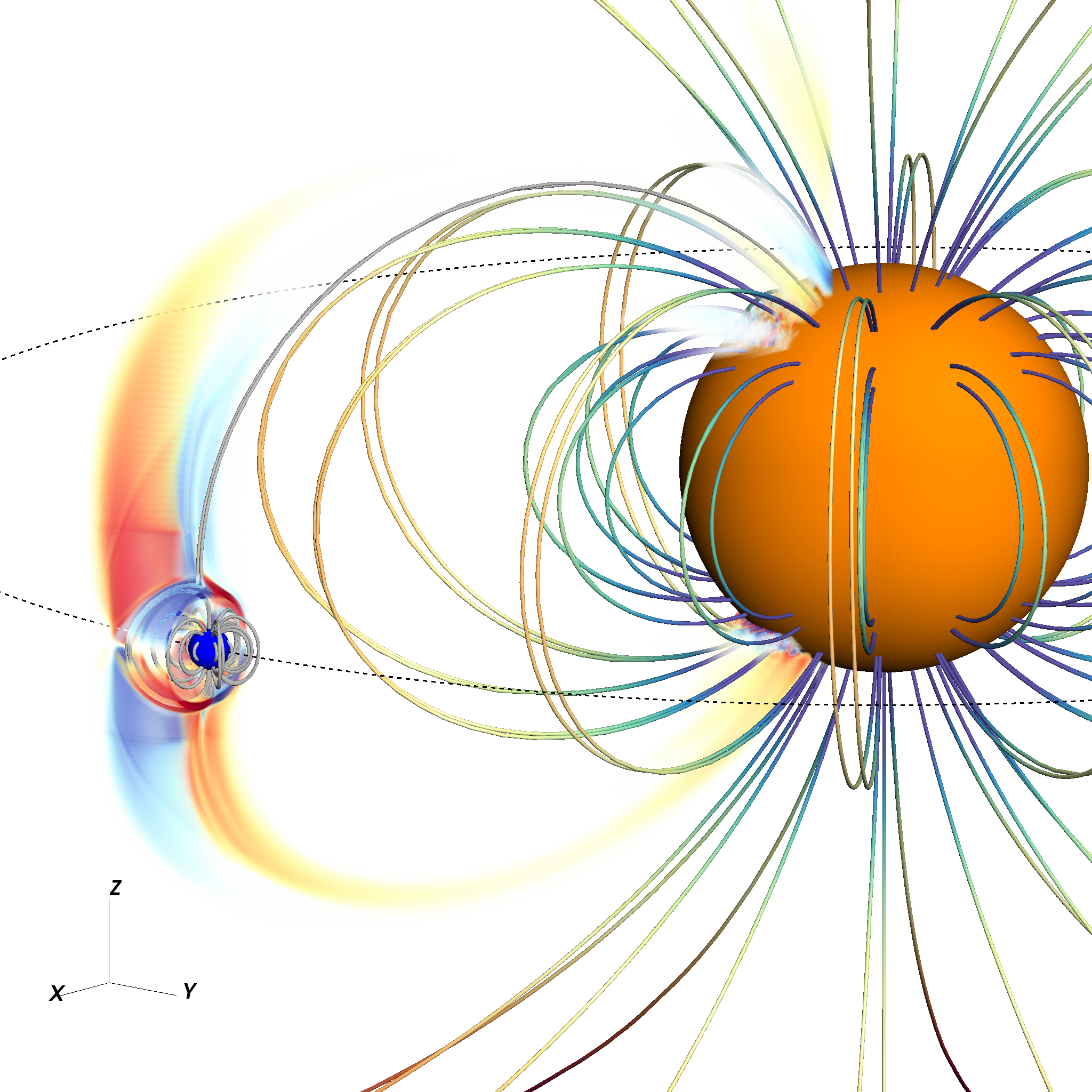}
  \includegraphics[width=0.4\linewidth]{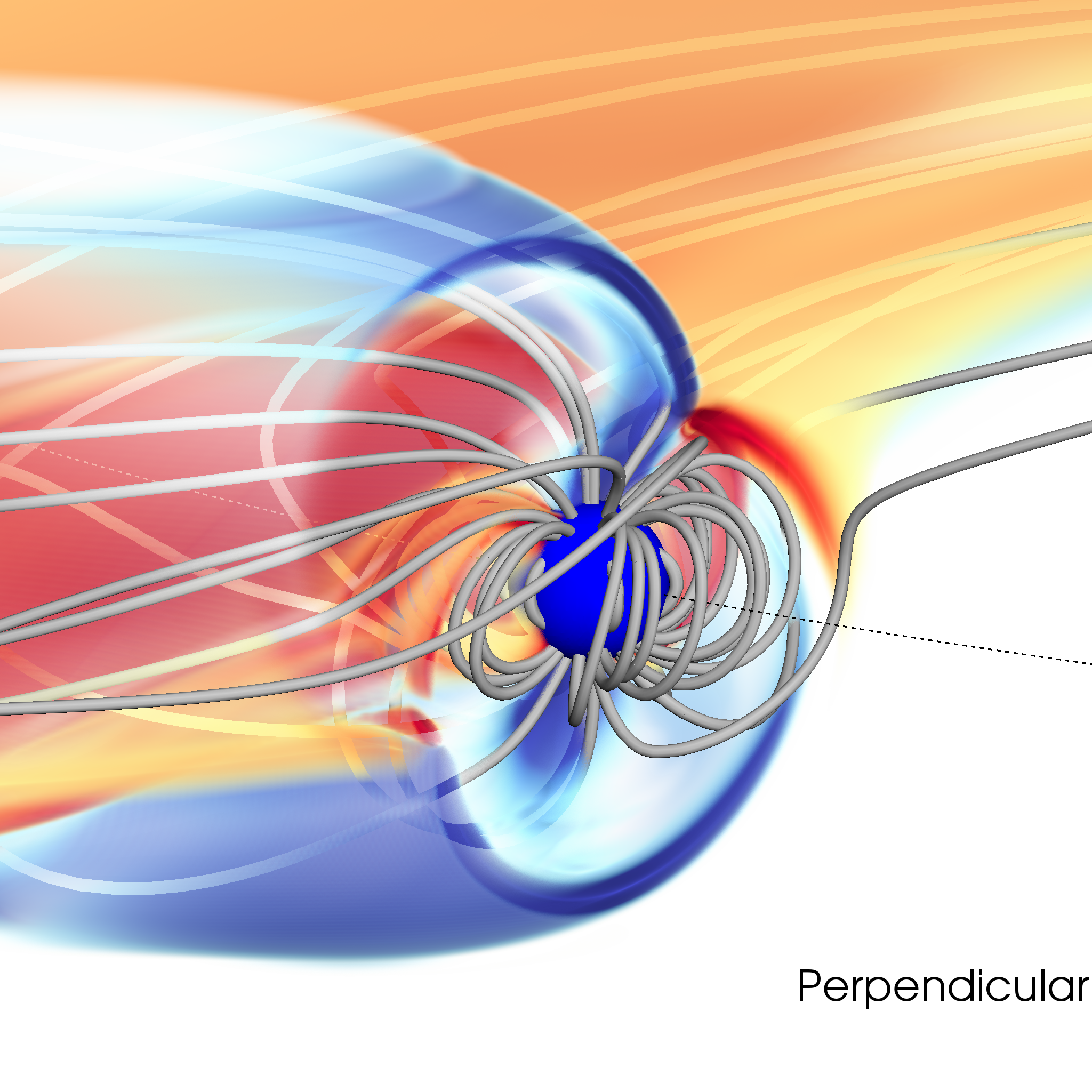}
  \includegraphics[width=0.4\linewidth]{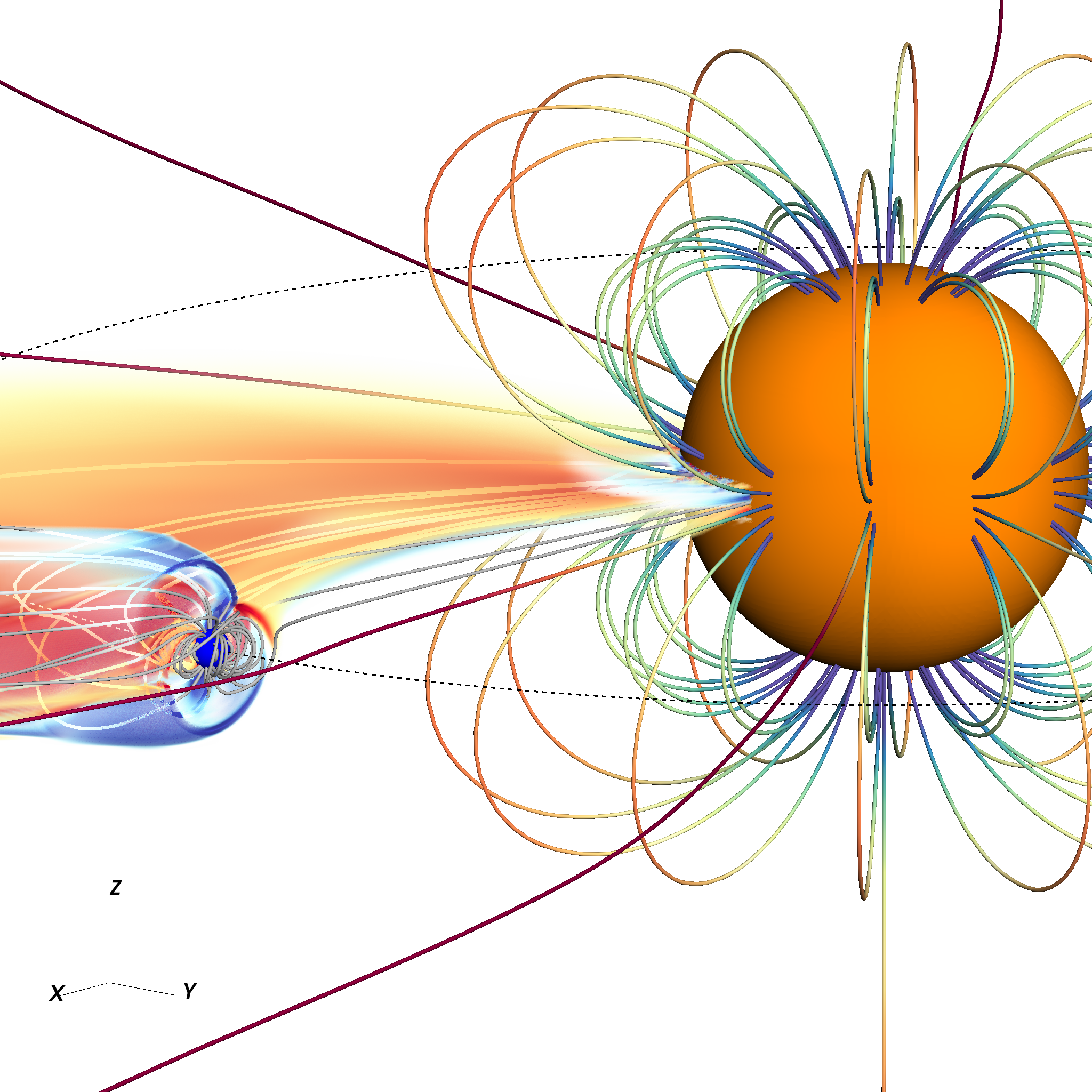}
  \caption{Three dimensional views of the aligned (top row),
    anti-aligned (middle row) and perpendicular (bottom row)
    configuration. The volume renderings represent the postive (red)
    and negative (blue) parallel currents (Equation \ref{eq:par_current}) delimiting the Alfv\'en
    wings. The volume is extruded from the star-planet plane to make
    its internal structure apparent. As a result the
    upstream-downstream asymmetry of the interaction is not visible,
    it will appear mode clearly in Figure \ref{fig:alfw_2d}. The stellar wind magnetic field
    lines are logarithmically
    color-coded with the magnetic field strength, and planetary magnetic field
    lines are shown in grey. The dashed black circle traces the orbit
    of the planet. The blue sphere represents the planet boundary, and
    the orange sphere the stellar boundary.}
  \label{fig:alfw_3d}
\end{figure*}

We display in Figure \ref{fig:alfw_3d} the global structure of the
Alfv\'en wings (right panels) with close-ups on the vicinity of the
planet (left panels) for each of three cases we consider. The
parallel currents 
\begin{equation}
  \label{eq:par_current}
  j_{||}={\bf J}\cdot\frac{{\bf B}}{\left|{\bf B}\right|}\, ,
\end{equation}
often referred to as \textit{Alfv\'en wing currents} \citep[due to the fact that
  they delimit Aflv\'en wings, see e.g.][]{Preusse:2006iu,Jia:2008fe},
  are shown by the red (positive) and
blue (negative) volume renderings. These volume rendering are extruded
at the vertical star-planet plane to make their internal structure
apparent. The magnetic field lines connected
to the planet are shown in gray, and the magnetic field lines
connected to the star are color-coded by the logarithm of the magnetic
field strength. The orbit of the planet is symbolized by the black
dashed circle, and the star and the planet are respectively
represented by the orange and blue spheres. 

\begin{figure*}[htb]
  \centering
  \includegraphics[width=\linewidth]{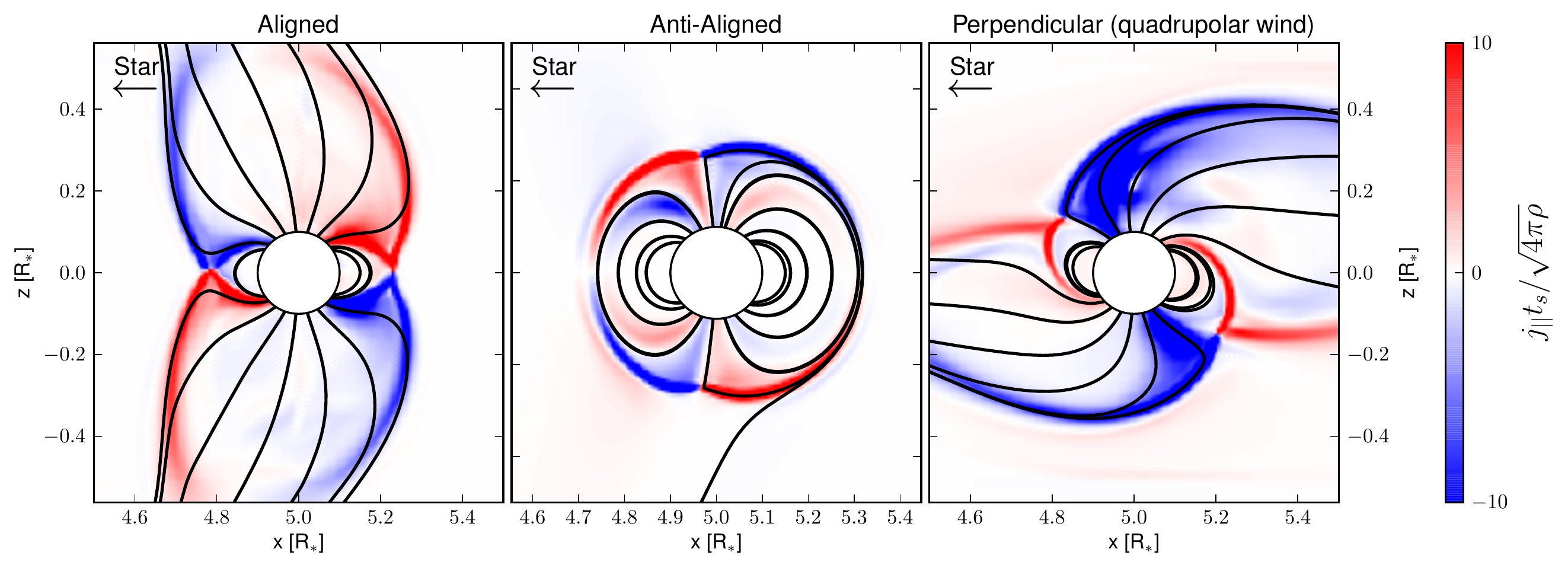}
  \includegraphics[width=\linewidth]{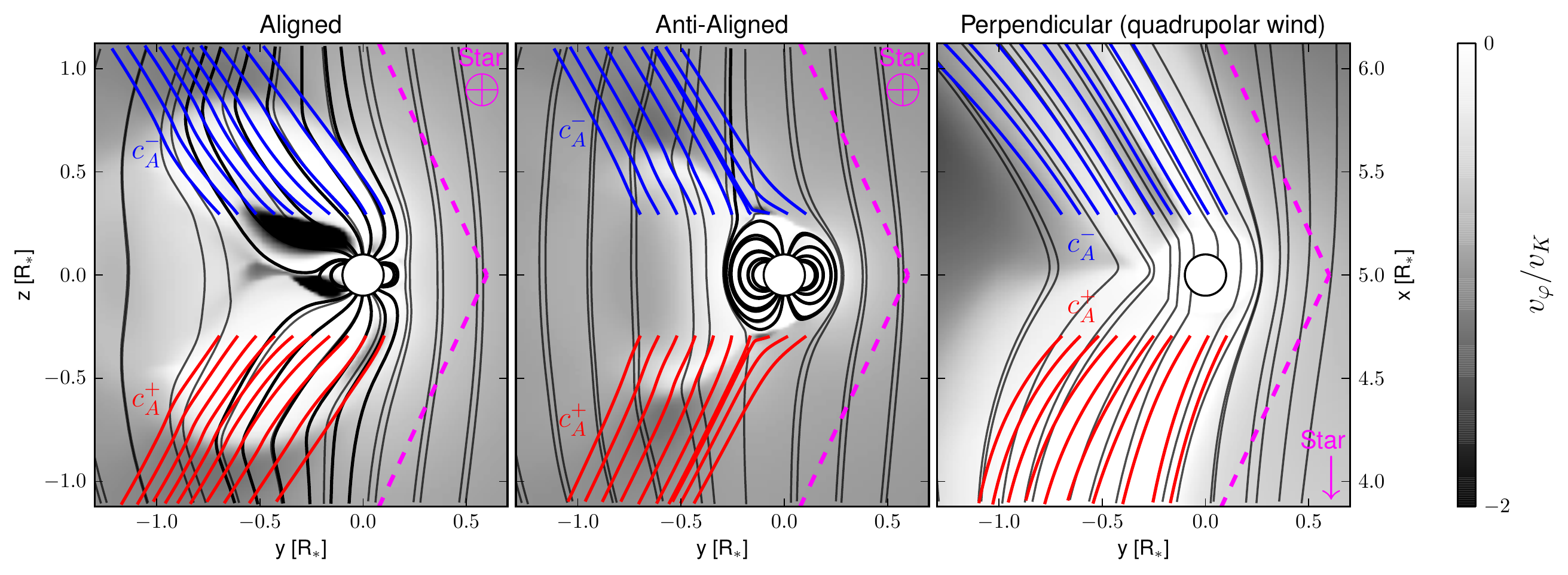}
  \includegraphics[width=\linewidth]{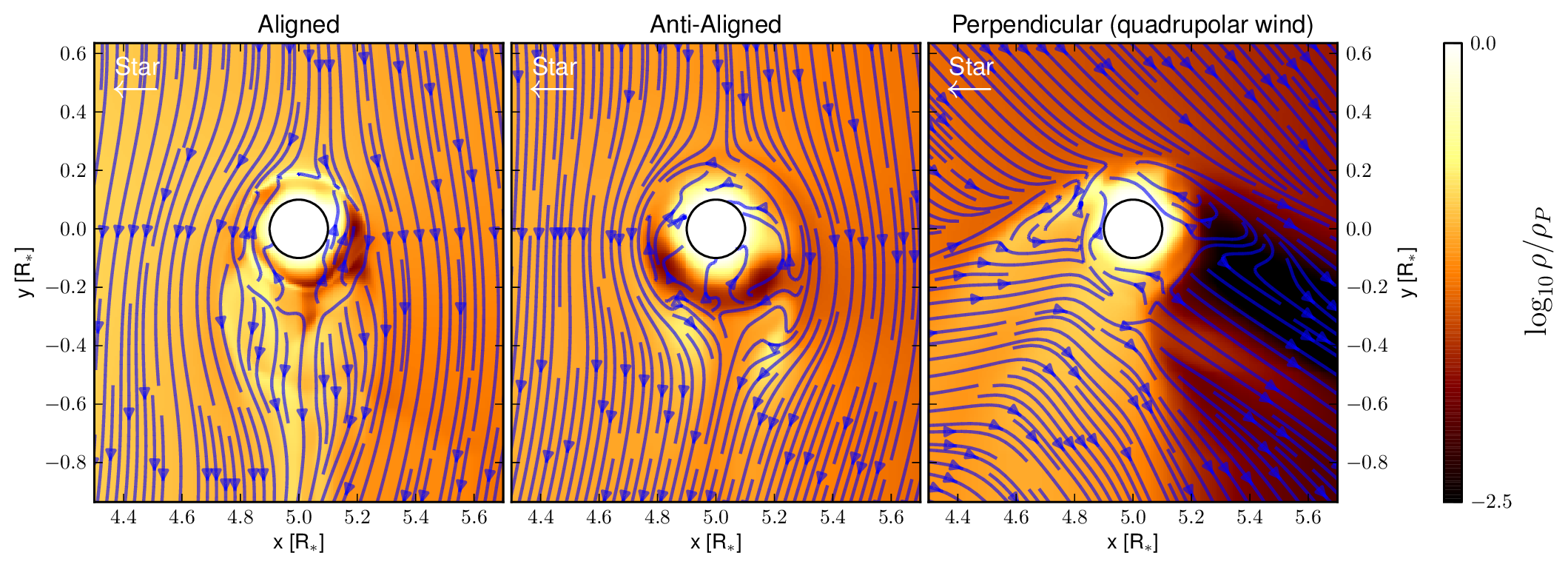}
  \caption{\textit{Top.} Positive (red) and
negative (blue) parallel currents in the star-planet plane
perpendicular to the orbital plane. The currents are normalized with the advection time-scale
across the planet (see text). The black lines represent the magnetic
field lines, and the white circle represents the planet.
\textit{Middle.} Cuts of the $({\bf v}_0,{\bf B}_w)$ plane. The
\reff{grey shades} shows the azimutal velocity in the rest frame of the planet,
normalized by the keplerian velocity $\bf{v}_K$.
The Alfv\'en characteristics
streamlines away from the planet are shown in blue and red. We do not
plot the Alfv\'en characteristics inside the planetary magnetosphere,
where they do not correspond to the travelling path of the
perturbations forming the Alfv\'en wings. The
expected inclination angle $\Theta_A^{\rm th}$ of the Alfv\'en wings is shown by the purple
dashed line. \textit{Bottom.} Density (in logarithmic
scale) in the orbital plane close to the planet. The streamlines of
the flow in the orbital plane are shown in blue.}
  \label{fig:alfw_2d}
\end{figure*}

In the case of a dipolar stellar magnetic field (upper and
middle panels),
the two Alfv\'en wings are connected to the star at high
latitude. Conversely, in the quadrupolar stellar wind (lower panels),
only one of the two Alfv\'en wings is connected to the star on the
magnetic equator. The footprint of the wings is 
generally out of phase from the planetary orbital phase due to the
finite propagation time of the alfv\'enic disturbances from the planet
vicinity to the stellar surface \reff{\citep[see][]{Preusse:2006iu,Kopp:2011if}}. In all cases the alfv\'enic
perturbations \reff{rapidly} travel the planet-star distance along the Alfv\'en wings
in less than 2 hours, which is less than $6\%$ of the orbital time
$t_{\rm orb}=2\pi R_{\rm orb}/v_K\sim 1.3$ days. \reff{As a result,
  we observe a small phase lag due to the fast propagation time and
  the small inclination angle of the wings (see hereafter Equation \ref{eq:angle_wing} and
  Table \ref{tab:tab0}).} The travel time is
nevertheless always larger than the typical advection time across the
planetary diameter $t_s=2R_P/v_K \sim 10$ minutes, due to the fast orbital motion of the
close-in planet. Hence, these perturbations are likely to never reach
back the orbiting planet and our simulations are always in 
the pure Alfv\'en wings regime. 
% The location of the footprint of the Alfv\'en wings on
% the stellar surface varies with the magnetic topology of the
% interaction, as seen on the right panels of Figure
% \ref{fig:alfw_3d}. In the aligned and anti-aligned cases the Alfv\'en wings clearly hit
% the stellar boundary at high latitude, whereas it impacts on the
% equatorial plane in the perpendicular case. The footprint of the
% unique Alfv\'en wing reaching back the star in this case
% is very elongated in longitude.
The upper and middle right panels
differ only by an inversion of the planetary magnetic field. We
immediately remark the importance of topology: in the aligned case the
Alfv\'en wings are broader and the currents much stronger
than in the anti-aligned case. As a result the Alfv\'en wings are
expected to carry much less energy and angular momentum in the anti-aligned case.

In the upper panel of Figure \ref{fig:alfw_2d} we quantify the
parallel currents nearby the planet. We show a side cut of the planet vicinity (the
star is located leftwards in those cuts), for
which the keplerian orbital motion of the planet is into the plane.
The black lines represent the magnetic field
lines. The parallel currents are again
shown by the red (positive) and blue (negative) colormap. They are
normalized by $t_s/\sqrt{4\pi\rho}$, thus effectively estimating
the ratio between $t_s$ and the time-scale associated with the
parallel currents. The most intense currents are
found at the boundary between the planetary magnetic field lines and the
wind: they originate from the planet-wind interaction as a simple
rotational discontinuity. They are associated with very short
time-scales compared to the advection time-scale $t_s$ and trace
the --comparatively-- fast reconnection rate between the planetary and wind magnetic
fields as the planet orbits in the wind. We stress here again that
the aligned and anti-aligned cases (two left panels) develop very different interaction
patterns. In the aligned case, the
strong currents are localized at the open--close magnetic field lines
boundary and at the Alfv\'en wing boundaries. In the anti-aligned case,
they are mainly localized at the magnetsophere-wind
boundary. The perpendicular case shows an in-between situation where
the strong currents delimit the Alfv\'en wings boundaries, and
the magnetosphere-wind interface.

The Alfv\'en wings plane $({\bf v}_0,\mathbf{B}_w)$ is shown in the
middle panels. The black lines again represent the
magnetic field lines, where the thicker lines are the planetary
magnetic field lines (the planetary magnetic field lines pervade the
Alfv\'en wings plane in the perpendicular case, since the Alfv\'en
wings plane coincided with the orbital plane in this case). In all cases the stellar
wind magnetic field lines are bent downstream due to the interaction
with the planet. The difference between aligned and anti-aligned
configurations clearly appears:
in the aligned case the extended polar magnetic field lines of the planet
allow for a large area of magnetic interaction with the wind, whereas
in the anti-aligned case the planetary magnetopshere remains in a
closed configuration. 
In both cases, though, Alfv\'en wings develop, symbolized by the blue
($c_A^-$) and red ($c_A^+$) streamlines of the Alfv\'en
characteristics. The expected theoretical inclination angle $\Theta_A$ between the ambient
magnetic field and the Alfv\'en wings is shown
by the magenta dashed lines and is given in those cases by \citep{Saur:2013aa}
\begin{equation}
  \label{eq:angle_wing}
  \sin\Theta_A^{\rm th} = \frac{M_A\sin\Theta}{\sqrt{1+M_A^2-2M_A\cos\Theta}}\, , 
\end{equation}
where the wind Aflv\'en Mach number is defined by $M_A=v_0/v_A$ 
and $\Theta$ is the angle between $v_0$ and ${\bf B}_w$ (note that in
\citealt{Saur:2013aa} $\Theta$ is the departure of perpendicularity
between ${\bf v}_0$ and ${\bf B}_w$). \reff{It is worth noticing that
  the theoretical inclination angle $\Theta_A^{\rm th}$ is expected to
be independant of the planetary magnetic field strength, which is held
fixed in our three cases.} The theoretical
estimate $\Theta_A^{\rm th}$ compares reasonably well with the simulated Alfv\'en
wings inclination angle $\Theta_A$ (averaged over the two wings, the characteristics of the
Alfv\'en wings are summarized in
Table \ref{tab:tab0}). The background colormap
shows the aziumthal velocity in the frame where the planet is at rest,
normalized to the Keplerian velocity $v_K$. In all cases
the magnetosphere of the planet orbits with the planet (white
regions), as well as the portion of the Alfv\'en wings intersected by
the cutting plane.

Finally, in the lower panels we display the flow in the rest frame of
the planet (blue streamlines) and the plasma density (logarithmic
colormap, normalized to the planet density) on the planetary orbital plane. In the first two panels
the flow is primarily in the orbital direction because the planet
orbits inside the dead-zone of the stellar corona, where the radial flow of
the stellar wind is negligible. The effective obstacle is 
larger on the equatorial plane in the anti-aligned case due
its larger magnetospheric extent (this is also apparent in the middle
panels). The effective obstacle is nevertheless three-dimensional and
is generally much bigger in the aligned case (see Section
\ref{sec:magnetic-torques}). No strong wake is observed downstream
since we chose to neglect planetary outflows, and that the interaction
is sub-alfv\'enic. In the third panel, the
stellar wind magnetic field lines are open in the
orbital plane. The flow is consequently composed of the orbital motion and the accelerating
radial wind. The planetary obstacle is observed to be much larger than
the planetary magnetosphere. Indeed, the Alfv\'en wings extend toward
and away from the star on the orbital plane and act as a supplementary
obstacle to the flow (see also the $v_\varphi$ colormaps in the middle
panels). Since we considered the
idealized case of a planet with no intrinsic mass loss, the high planet density
does not propagate to more than a few grid points inside the planet
magnetosphere. Varying the planet internal density only marginally
affects our results. The detailed density pattern in the planetary
magnetosphere is found to have very little impact on the
Alfv\'en wings properties.

\begin{deluxetable}{lccc}
  \tablecaption{Properties of Alfv\'en wings\label{tab:tab0}}
  \tablecomments{The theoretical $^{\rm th}$ values are estimates from
    the analytical model of \citet{Saur:2013aa} (in which the
    anti-aligned case is not modelled).
  }
  \tablecolumns{4}
  \tabletypesize{\scriptsize}
  \tablehead{
    \colhead{} &
    \colhead{Aligned} &
    \colhead{Anti-aligned} & 
    \colhead{Perpendicular}
  }
  \startdata 
  $\Theta_A^{\rm th}$ [$^\circ$] & 25 & [-] & 25 \\
  $\Theta_A$ [$^\circ$] & 29.7 & 28.2 & 27.3 \\
  $R_{\rm eff}^{\rm th}$ [$R_P$] & 3.6 & [-] & 3.0 \\
  $R_{\rm eff}$ [$R_P$] & 3.0 & 1.2 & 2.2  \\
  $\mathcal{P}^{\rm th}$ [W] & 2.05 10$^{19}$ & [-] & 1.09 10$^{18}$\\
  $\mathcal{P}$ [W] & 1.39 10$^{19}$ & 9.74 10$^{17}$ & 7.72 10$^{17}$
  \enddata
\end{deluxetable}

\subsection{Poynting flux in Alfv\'en wings}
\label{sec:poynting-flux-afven}

\begin{figure*}[htb]
  \centering
  \includegraphics[width=\linewidth]{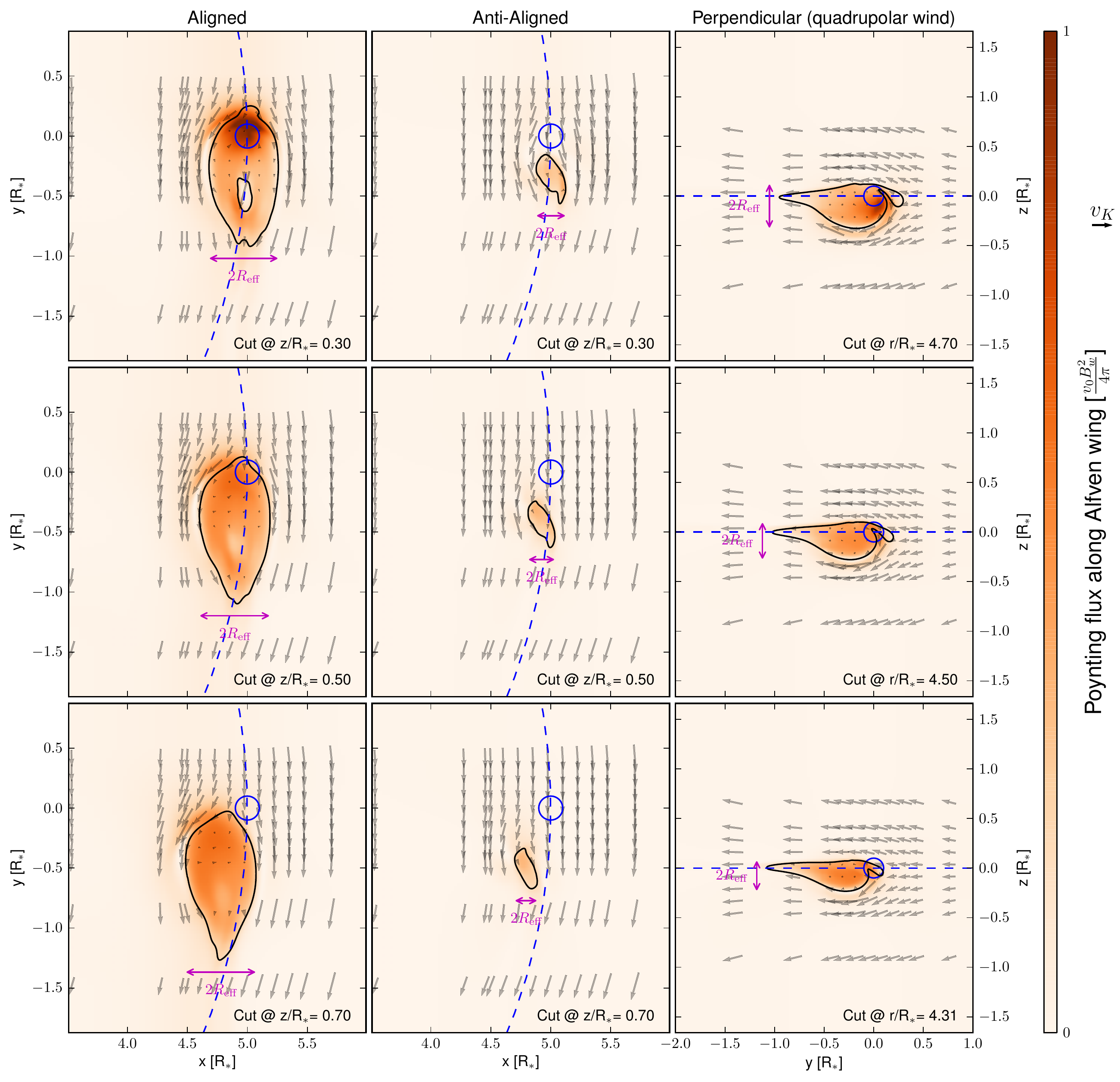}
  \caption{Poynting flux along Alfv\'en wings. The rows correspond to
  cuts of the Alfv\'en wing at distances of 0.3, 0.5 and 0.7 $R_\star$
  from the planet. The Poynting flux is normalized to the maximum
  expected Poynting flux $v_0B_w^2/4\pi$. The black contours
  represent the boundary of the Alfv\'en wing, identified as the
  region co-orbiting with the planet in the projection plane. The
  Alfv\'en wing cross-section perpendicular to the flow is indicated in
  purple. The black arrows represent the
  flow on the cutting plane in the rest frame of the planet. The dashed blue line
  is the projection of the orbital trajectory, and the blue circle the
  projection of the planet boundary.}
  \label{fig:poynting_flux}
\end{figure*}

The magnetic interaction is a source of magnetic energy transfer
between the wind, the planet and the host star. It could be a source
for observable emissions, its characterization is thus of
major importance for the search for exoplanets today. The
Poynting flux in each Alfv\'en wing can be evaluated by
\begin{equation}
  \label{eq:Poynting_flux}
  S_a = \frac{c{\bf E}\times{\bf B}}{4\pi} \cdot \frac{{\bf c}_A^\pm}{|{\bf c}_A^\pm|}\, ,
\end{equation}
where the electric field is \reff{$c{\bf E}
= - {\bf v}\times{\bf B}$} in the ideal MHD approximation. The Poynting flux depends on the
frame in which it is calculated, as a result we consider here the \reff{inertial} reference frame
to mimic what a distant observer would see when observing such a
system. Because the central star slowly rotates, this frame also
conveniently corresponds to the stellar reference frame, and the
Poynting flux corresponds to the energy that may be
deposited on the star due to the SPMI. We display in Figure \ref{fig:poynting_flux} the Poynting
flux on horizontal cutting planes along the $c_A^-$ Alfv\'en wing for the aligned and
anti-aligned cases, slightly above the equatorial plane at
$z=0.3,\, 0.5,\, 0.7\,R_\star$. For the quadrupolar stellar wind (right panels) the
Alfv\'en wings are centered on the orbital plane, hence we display the
Poynting flux along the $c_A^+$ Alfv\'en wing on vertical cutting
planes in between the planet and the star. In each panel,
the blue circle and the dashed line respectively represent the projection of the
planet and of its orbital trajectory on the cutting plane. 

Comparing the
aligned and anti-aligned cases (left two panels), we immediately observe the strong
Poynting flux concentrated inside the Alfv\'en wing in the 
aligned case, while in the anti-aligned case the Poynting flux is
extremely weak. In all cases the flux is nevertheless positive,
denoting a flux of energy towards the star. In the aligned and perpendicular cases
the Alfv\'en wing is tear-shaped in the direction of the flow (grey arrows), and the
maximum Poynting flux is localized close to the side of the wing
facing the flow. In these two cases the Alfv\'en wing clearly acts as an obstacle
to the flow. Conversely, in the anti-aligned case, the very small cross-section
of the Alfv\'en wing makes only a small perturbation to the flow, as
seen in the middle panels. As the cutting plane is shifted away from
the planet (from top to bottom), the Alfv\'en wing center is observed
to shift towards the star and downstream. 

The cross-section of the Alfv\'en wing perpendicular to the flow,
$2R_{\rm eff}$, is indicated by the double arrows in Figure
\ref{fig:poynting_flux}, and computed by taking the maximal extent of
the wing perpendicular to the flow. It remains approximately constant along the wing as long as
the cutting plane remains roughly perpendicular. \citet{Saur:2013aa}
used a simple magneto-static equilibrium
code to estimate the expected $R_{\rm eff}$ as a function of the standard
obstacle radius
$R_{\rm obst}$ \citep[\textit{e.g.},][]{Lovelace:2008bl,Lanza:2009fp}
which are defined by
\begin{eqnarray}
  \label{eq:reff}
  R_{\rm eff} &\sim& R_{\rm obst}
                     \left(3\cos\frac{\Theta_M}{2}\right)^{1/2}
                     \, ,\\
  R_{\rm obst} &\sim& R_p\left(\frac{B_P^2}{8\pi P_t}\right)^{1/6}\, ,
  %R_{\rm obst} &\sim& \left(\frac{2\mu_P^2}{P_t}\right)^{1/6}\, ,
  \label{eq:robst}
\end{eqnarray}
where $P_t$ is the
total pressure in the vicinity of the planet, and $\Theta_M$ is the
inclination angle between the polar planetary field and ${\bf
  B}_w$ \reff{\citep[we focus here on the case of a
  magnetized planet, the effective obstacle in the case of a planet
  with no intrinsic magnetic field is discussed, \textit{e.g.},][]{Kopp:2011if}}. Because of the magneto-static equilibrium
assumed to derive
Equation \eqref{eq:reff}, we do not expect it to exactly match our
observed effective radii $R_{\rm eff}$. The
estimated and simulated effective radii $R_{\rm eff}$ are
given in Table \ref{tab:tab0}. It appears that the theoretical
value slightly over-estimates the effective radius we obtain in our
simulation. The shape of the simulated Alfv\'en wing
cross-section is much more elongated --in the flow direction-- than
the theoretical Alfv\'en
wing of \citet{Saur:2013aa}. As a result, this effect compensates the
discrepancy in the effective radii of the obstacle, and the
theoretical and simulated wings have similar cross-section areas.

The maximum amplitude of
the Poynting flux scales remarkably well with the predicted value of
$v_0B_w^2/4\pi$ expected from the analytical estimates of 
\citet{Saur:2013aa}. By integrating the total
Poynting flux inside the area $\Sigma_A$ delimited by black contours
in Figure \ref{fig:poynting_flux}
(effectively delimiting the Alfv\'en wing cross-section), we find that
the total Poynting flux is
close to being constant throughout each wing. We report the
average value of the integrated Poynting flux
\begin{equation}
  \label{eq:integral_poynting_flux}
  \mathcal{P} = \left\langle \int S_A\,  {\rm d}\Sigma_A \right\rangle_A
\end{equation}
in Table \ref{tab:tab0}, where $\left\langle \right\rangle_A$ stands
for an average along the Alfv\'en wing. The theoretical and simulated
Poynting fluxes agree within a factor of two, which is satisfying
given the approximations embedded in both the analytical and
numerical models. The total Poynting flux is smaller by a factor of
$6$ between the aligned and the anti-aligned cases. This
topological effect could provide a simple explanation to the on/off
enhanced emissions observed in extreme exo-systems over the
time-scales of typical stellar magnetic cycles or orbital periods.

\section{Planet migration due to magnetic torques}
\label{sec:magnetic-torques}

We have demonstrated that our numerical model is able to simulate
adequately Alfv\'en wings that compare well with analytical
theory estimates. We now focus on the less studied magnetic torques,
which can play a role in the secular evolution of close-in star-planet
systems.

\subsection{Physical origin of the torques}
\label{sec:origin-torques}

Torques in magnetic star-planet systems can be separated in
contributions from ram pressure, thermal
pressure, magnetic pressure, and magnetic tension. We derive the angular
momentum fluxes associated with those torques in a frame rotating at the
planetary orbital rate. The angular momentum is defined with respect
to the stellar rotation axis, which coincides with the orbital axis for
the cases considered in this work. The interested reader will find the derivation
of the various torques expressions in Appendix \ref{sec:expr-torq-star}.

\begin{figure*}[htb]
  \centering
  \includegraphics[width=\linewidth]{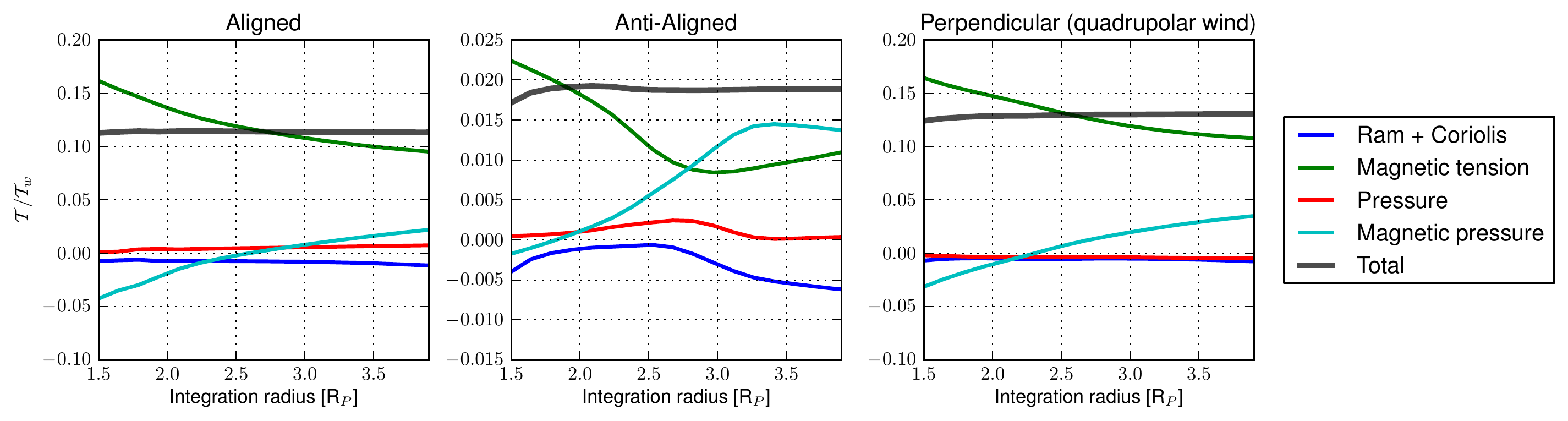}
  \caption{Torques applied to the planet integrated over concentric
    spheres around the planet. The torques are normalized to the
    stellar wind torque in each case. They are separated into
    contributions from the ram pressure and Coriolis force, thermal
    pressure, magnetic pressure, and magnetic tension (see Appendix
    \,\ref{sec:expr-torq-star} for details). The total torque is
    indicated in black.}
  \label{fig:torq_p}
\end{figure*}

Thanks to angular momentum conservation, these torques can be
evaluated on any surface enclosing the planet. As a result we estimate
the torques from integrations over concentric spheres around the
planet, and \textit{a posteriori} check that the total 
torque $\mathcal{T}$ (black lines in Figure \ref{fig:torq_p}) is indeed constant with the
integration radius. The different components of the torque are detailed
in Figure \ref{fig:torq_p}.

In the aligned and perpendicular cases (left and right panels), the torque is dominated by the
tension of the magnetic field lines connecting the planet to the star
(green line). It is slightly opposed by the magnetic pressure (cyan
line) while the ram and thermal pressures play almost no role in the
the overall torque. Note that in the three cases, the magnetic pressure
(cyan) changes sign as the integration sphere surrounding the planet is
expanded. This is a simple consequence of the fact that the
magnetosphere of the planet is more extended downstream than
upstream. As a result, the integration is
imbalanced toward the downstream contribution and gives a negative contribution when the
integration spheres are fully inside the planetary magnetosphere. When the
integration sphere is fully outside the planetary magnetosphere, the
magnetic pressure contribution to the torque is positive, as
expected. 

In the anti-aligned case (middle panel),
the total torque is the combination of both magnetic tension and
pressure, corresponding to the wind magnetic field lines
impacting the closed planetary magnetosphere. The ram and
thermal pressures play here a marginal role as well in the
total torque. The torque applied to the planet in the aligned case is
roughly $10\%$ of the stellar wind torque applying to the host
star. Because the two Alfv\'en wings connect back to the star, it
means that the star brakes $10\%$ less efficiently than a
twin star not harbouring any close-in planet.
The torque in the aligned case is furthermore approximately five times
higher than in the
anti-aligned case, showing again the very strong impact of the
magnetic topology on the strength of the SPMI. The perpendicular case
shows a very similar repartition of the different contribution than
the aligned cases, with a slightly higher magnetic pressure. It is
interesting to note that in the perpendicular case only one wing
connects the star and the planet, as a result only half of the angular
momentum extracted from the orbit of the planet will be transfered to
the star, while the other half will be advected in the wind. Hence,
the impact of the SPMI on the host star is maximized in the
aligned case.

We further calculate the migration time-scale
associated with those torques, given by
\begin{equation}
  \label{eq:mig_timescale}
  t_P = \frac{2J_P}{\mathcal{T}}\, ,
\end{equation}
where $J_P=M_P\left(GM_\star R_{\rm orb}\right)^{1/2}$ is the orbital angular
momentum of the planet\reff{, and the factor $2$ originates from the
  $R_{\rm orb}^{1/2}$ dependance of $J_P$}. The migration time-scale depends on the
density normalization $\rho_\star$. In Table \ref{tab:tab1} we give the
migration timescales (as well as the average torques $\mathcal{T}$)
for a $\rho_\star$ normalization corresponding
to a TTauri-star mass-loss rate (five orders of magnitude higher than
the Sun, corresponding as well to a much stronger stellar magnetic
field). The time-scales are clearly sufficiently short to
suggest that magnetic interactions can play a role in the migration of
close-in planets, especially during the early stages (typically TTauri
and pre-main sequence phases) of star-planet
systems. The density normalization divides the migration time-scale
$t_P$ when computing it from adimensionalized units. For a solar-like
density normalization (see Section \ref{sec:stellar-winds}), the
time-scales would be 5 orders of magnitude higher and be negligible
compared to typical tidal effects. Preliminary scaling-laws for the
migration timescales of close-in planets were derived
\citet{Strugarek:2014ab} from reduced 2.5D simulations. The variation
of $t_P$ with topology in our 3D models agrees with the predicted law from
\citet{Strugarek:2014ab}, and the orbital radius dependency is not
expected to significantly change from 2.5D to 3D models. Though,
the torques differ qualitatively
because the multiplicative constant in front of the scaling law
derived in \citet{Strugarek:2014ab} was calibrated with 2.5D models.
By using a grid of 3D numerical simulations (currently under
investigation), we intend to better constrain this multiplicative
constant to obtain a quantitatively accurate scaling law in a near
future. This extended set of simulations shall also confirm the
dependency of the migration timescale $t_P$ on the orbital radius
$R_{\rm orb}$ found in \citet{Strugarek:2014ab}. Finally, close-in
star-planet systems can also be in a
super-alfv\'enic interaction regime (not explored here). Indeed, even
if the wind in the vicinity of the planet is sub-alfv\'enic, the
relative fast orbital motion can exceed the local Alfv\'en speed. How the
torque scaling-law changes between the sub- and super-alv\'enic regimes
still remains to be explored.

\subsection{Parametrization of magnetic torques and effects of
  magnetic topology}
\label{sec:param-magn-torq}

By analogy with an obstacle in a flow, the magnetic
torque applied to the planet due to the SPMI
is generally written as \citep[\textit{e.g.},][]{Lovelace:2008bl,Vidotto:2009hb}

\begin{equation}
  \label{eq:proto_torq}
  \mathcal{T} = c_d  \,  R_{\rm orb} \, A_{\rm obst}\, P_t \, ,
\end{equation}

where $A_{\rm obst}$ is the effective obstacle area exposed to the
flow, $P_t$ the total (thermal plus ram plus magnetic) pressure of the wind in the
frame where the planet is at rest, and
$c_d$ a drag coefficient. The right hand side is conveniently composed of
the total angular momentum that can be transfered, multiplied by $c_d$.
In the case of SPMI, the drag coefficient $c_d$ and the effective area
$A_{\rm obst}$ should generally depend on the
topology of the interaction, \textit{i.e.} on the
respective orientations of the orbital motion, the interplanetary
magnetic field, and the planetary magnetic field. Due to this
complexity, the drag coefficient $c_d$ and the effective interaction
area $A_{\rm obst}$ can be non trivial to estimate.

The drag coefficient is generally thought to represent --in the case
of SPMI-- the reconnection efficiency between the stellar wind
and the planetary magnetic fields, at the boundaries of the
planetary magnetosphere or of the Alfv\'en wings themselves. 
In the context of planetary radio emissions,
\citet{Zarka:2007fo} approximated $c_d$ with \citep[see also][]{Neubauer:1998aa,Saur:2013aa}
\begin{equation}
  \label{eq:reconnection_coeff}
  %c_d \sim \frac{2}{\sqrt{1+M_A^{-2}}}\, ,
  c_d \sim \frac{M_A}{\sqrt{1+M_A^2-2M_A\cos\Theta}}\, .
\end{equation}
% where $M_A$ is the local alfv\'enic Mach number near the obstacle.
This latter equation is not thought to be valid in the closed
magnetosphere case (here the so-called 'anti-aligned' case), for which
we simply choose $c_d\sim 1$. The drag coefficient in the aligned and
perpendicular cases is given in Table \ref{tab:tab1}.

\begin{deluxetable}{lccc}
  \tablecaption{Torques and effective obstacle areas of the magnetic interaction\label{tab:tab1}}
  \tablecomments{The theoretical area $A^{\rm th}_{\rm obst}$ is
    obtained with the expected magnetospheric size $R^{\rm th}_{\rm
      obst}$ from Equation \ref{eq:robst}. The migration time-scale
    $t_P$ is calculated with a base density $\rho_\star=3.2\,10^{-10}$
    g/cm$^3$ for the aligned and anti-aligned cases, and
    $\rho_\star=2.8\,10^{-11}$ g/cm$^3$ for the perpendicular case
    (see text).} 
  \tablecolumns{4}
  \tabletypesize{\scriptsize}
  \tablehead{
    \colhead{} &
    \colhead{Aligned} &
    \colhead{Anti-aligned} & 
    \colhead{Perpendicular}
  }
  \startdata 
  $\mathcal{T}$ [$\mathcal{T}_w$] & 0.11 & 0.02 & 0.13 \\
  $c_d$ & 0.43 & {\it 1.0} & 0.64 \\
  $A_{\rm obst}^{\rm th}$ [$\pi R_p^2$] & 4.4 & 4.4 & 4.0 \\
  %$A_{\rm eff}$ [$\pi R_p^2$] & 36.6 & 6.0 & 12.6 \\
  $A_{\rm obst}$ [$\pi R_p^2$] & 72.4 & 5.1 & 25.0 \\
  $t_P$ [Myr]  & 1.39\, 10$^2$ & 8.46\, 10$^2$ & 2.70\, 10$^3$
  \enddata
\end{deluxetable}

The obstacle area $A_{\rm obst}$ is generally considered as a circular
cross-section of the planetary magnetosphere,
of estimated radius $R_{\rm obst}$ (equation \ref{eq:robst}),
itself deduced from a simple pressure balance.
Though, it is not often recognized that this effective area
changes drastically with the magnetic topology and is, in general, far
from being
circular. We use here our numerical simulations to estimate $A_{\rm
  obst}$, based on the integrated torque $\mathcal{T}$ shown in Figure \ref{fig:torq_p}.
The resulting areas are given in Table
\ref{tab:tab1}, along with the standard obstacle area $A_{\rm
  obst}^{\rm th} = \pi R_{\rm obst}^2$. The anti-aligned case 
(middle panels in Figure \ref{fig:alfw_2d}) is the only situation in which the effective
obstacle is indeed the roughly circular magnetospheric
cross-section perpendicular to the flow (see also schematic in Figure
\ref{fig:final_schematic}). 

In the aligned and perpendicular cases
(upper and lower panels in Figure \ref{fig:alfw_2d}), the
connection between the planetary field and the wind magnetic field
leads to an interaction cross-section composed of the whole flux-tube
connecting the star and the planet (which width is given by $R_{\rm
  eff}$, see Table \ref{tab:tab0}), and hence to a
much greater torque (as seen in Figure \ref{fig:torq_p}). The
corresponding obstacle area is found to be 14 times higher than the
standard obstacle area in the aligned case, and 5 times higher in the
perpendicular case (note that the corresponding torques in Table
\ref{tab:tab1} are normalized to the stellar wind torques
$\mathcal{T}_w$, which differ in the dipolar and quadrupolar cases as
seen in Table \ref{tab:tabw}). The torque is maximized and
minimized in the two extreme cases of aligned and anti-aligned
topologies. As a result, since we
generally cannot infer the magnetic field of known exoplanets, these
two cases give good upper and lower estimates of magnetic
torques a given star-planet system can develop.

\section{Conclusions}
\label{sec:conclusions}

In this work we have simulated in three dimensions the magnetic interactions of
a star with a close-in planet. By simulating the system globally, we
were able to trace Alfv\'en wings extending from the planet
magnetosphere to the stellar lower corona. We have explored three
typical magnetic configurations of SPMIs: aligned, anti-aligned,
and perpendicular orientations of the planetary field with respect to
the ambient wind magnetic field. For the perpendicular case we chose
to consider a dipolar planetary field perpendicular to the orbital plane
and a quadrupolar stellar wind. In this latter case the accelerating
wind participates to the interaction and the Alfv\'en wings extend
near the orbital plane. In the aligned and anti-aligned cases, the planet
orbits inside a wind 'dead-zone' and the Alfv\'en wings extend out of
the orbital plane along the dipolar structure of the stellar corona.

The Poynting flux in Alfv\'en
wings provides an energy source for enhanced X or UV emissions in the
star-planet system. We were able to validate our numerical model by
comparing the simulated Poynting
fluxes with analytical predictions. We find
that the size of the Alfv\'en wings (and their associated Poynting flux)
dramatically depends on the magnetic configuration of the interaction:
by reversing the planetary field, the Poynting flux drops
approximately by a factor of 14. 

\begin{figure}[htb]
  \centering
  \includegraphics[width=\linewidth]{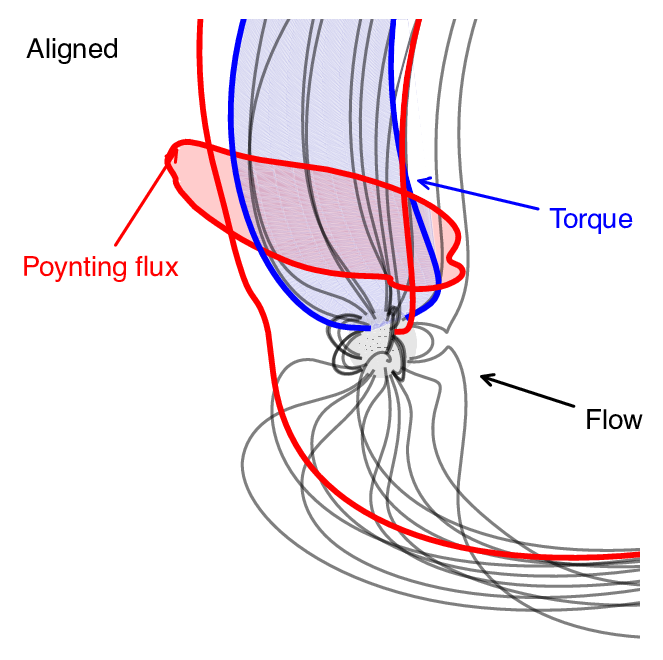}
  \includegraphics[width=\linewidth]{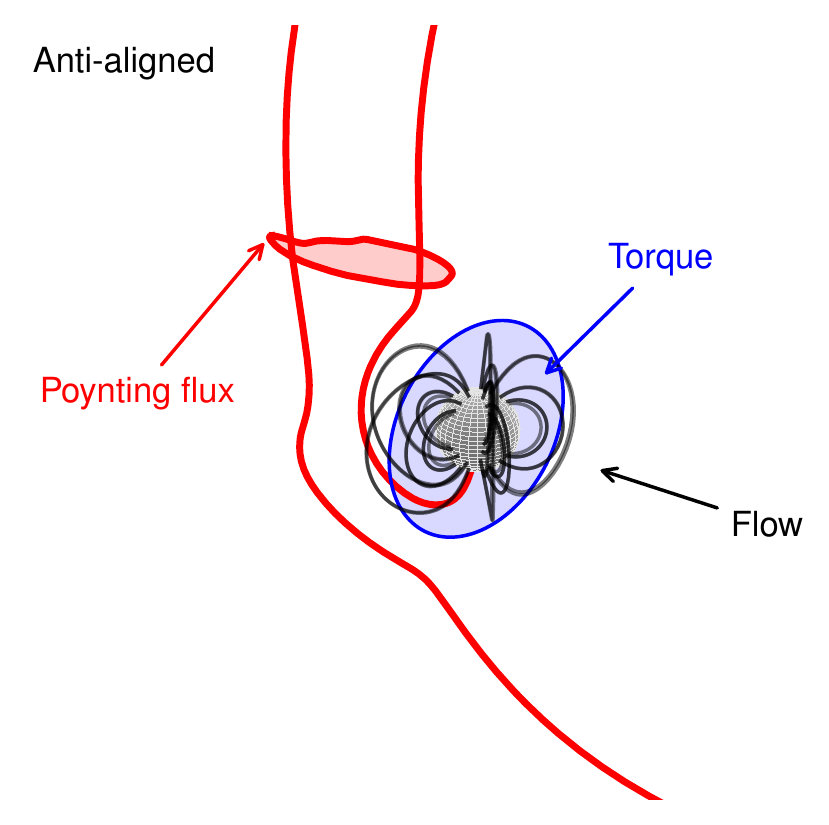}
  \caption{Schematics of magnetic star-planet interactions in the
    aligned (top) and anti-aligned (bottom) cases. The black lines
    represent the magnetic field lines, the blue and red lines delimit
    the upper Alfv\'en wing. Characteristic surfaces associated with
    the Poynting flux (red area) and the magnetic torques (blue area)
    are highlighted in each configuration, showing the critical role
    of magnetic topology in the development of the interaction.}
  \label{fig:final_schematic}
\end{figure}

We used numerical simulations to estimate the magnetic torques that
develop in star-planet magnetic interactions. Again, the magnetic
configuration affects significantly the torques that develop in such
systems. In the aligned case, a part of the planetary magnetosphere is
open in the ambient stellar wind and connects the star and the planet
together. They are a source of magnetic tension that effectively
transfers angular momentum between the star and the planet. In the
cases presented in this paper, the planet loses orbital angular
momentum and migrates inward \citep[for fast rotators a planet may
\textit{a priori} gain orbital angular momentum from such interaction,
for a detailed discussion see][]{Strugarek:2014ab}. In the aligned
case the migration time-scale of the studied
planets vary from 100 to 1000 Myrs for a TTauri-like host star. As
a result, magnetic interactions can be an important factor of migration
for young close-in planets. In the anti-aligned case the magnetosphere
is completely closed: the torque then originates from the impacting
coronal plasma on the cross-section of the magnetosphere. The area of
interaction is hence much smaller than in the aligned case, and the
associated magnetic torque is roughly 14 times smaller.

We illustrate the importance of magnetic configuration in Figure
\ref{fig:final_schematic} where the aligned and anti-aligned
configurations are schematized. The
red areas show the cross-section of one Alfv\'en wing that carries the induced
Poynting flux. The blue areas represent the interaction
areas leading to magnetic torques. The standard estimation of magnetic
torques, based on Equation \eqref{eq:proto_torq}, generally makes use of the effective
area of the anti-aligned interaction. We find here that this area
minimizes the magnetic torque that can develop in close-in star-planet
systems, while the blue area of the aligned case maximizes it.

The three-dimensional simulations reported in this work confirm the
2.5D axisymmetric results of \citet{Strugarek:2014ab}: magnetic
torques can be a source of close-in planet migration. We have focused
on characterizing the impact of magnetic configuration on the
shape and strength of the magnetic interactions. We are currently running
a more extensive set of models to empirically refine the scaling laws
first derived in \citet{Strugarek:2014ab}.

Real stars posses much more complex magnetic fields than the simple
dipolar and quadrupolar configurations we considered in this work. In
reality close-in planets are likely to interact with different local
magnetic configuration along their orbit \citep[see,
\textit{e.g.}][]{Cohen:2014eb,Strugarek:2014sa,Vidotto:2015hw}. Our
results suggest that, in such systems, the
associated Poynting fluxes and torques will vary by at least an order
of magnitude, which provides a simple geometrical
explanation for an on/off mechanism of magnetically enhanced
emissions in close-in star-planet systems. The
average Poynting flux and torque such systems develop are nonetheless non-trivial
to estimate. They will require dedicated 3D simulations tackling the
dynamical aspects of magnetic interactions as a planet orbits in a
non-homogenous corona. Indeed, the time-scale on which the equilibrated
configurations modelled in this paper establish depends on the
resistivity of the magnetospheric plasma of the planet, and on its
reconnection efficiency with the stellar wind magnetic field. The numerical
model presented in this work provides a solid basis for further, more realistic studies of
star-planet magnetic interactions in which these dynamical aspects
could be explored.

\acknowledgments

A. Strugarek is a National Postdoctoral Fellow at the Canadian Institute of
Theoretical Astrophysics, and
acknowledges support from the Canada’s Natural Sciences and
Engineering Research Council. This work was supported by
the ANR 2011 Blanc 
\href{http://ipag.osug.fr/Anr\_Toupies/}{Toupies}
and the ERC project \href{http://www.stars2.eu/}{STARS2}. We acknowledge access to supercomputers
through GENCI (project 1623), Prace (8th call), and ComputeCanada infrastructures.

\bibliographystyle{yahapj}
%\bibliography{/Users/astrugar/WORK/MyPapers/mybib}

\appendix

\section{General expressions of torques}
\label{sec:expr-torq-star}

We derive here the general expression of torques in a rotating frame,
based on the initial derivation of \citet{1970MNRAS.149..197M}.
The momentum equation in the MHD formalism in a rotating frame (with a
rotation rate $\boldsymbol{\Omega}$) can be written
\begin{equation}
  \label{eq:mom}
  \partial_{t} \left(\rho\mathbf{u}\right) = \Div\left( -\rho\mathbf{u}\mathbf{u} +
    \mathbf{B}\mathbf{B}/4\pi - \mathbf{I}P_{t}\right) + \rho\mathbf{g} -
  2\rho\boldsymbol{\Omega}\times\mathbf{u} + \rho\boldsymbol{\Omega}\times\boldsymbol{\Omega}\times\mathbf{r}\,,
\end{equation}
where the total pressure is the sum of the thermal pressure and the
magnetic pressure $P_{t} = P + B^{2}/8\pi$. We define the vectorial angular momentum by
\begin{equation}
  \label{eq:amom_def}
  \boldsymbol{\mathcal{J}} \equiv \int_{V}\mathbf{r}\times \rho\mathbf{u} \,{\rm d}V\, ,
\end{equation}
with $V$ a given volume. The time evolution of the angular momentum, or equivalently the
torques acting on the volume $V$, is given by
\begin{equation}
  \label{eq:jdot}
  \dot{\boldsymbol{\mathcal{J}}} =
  \int_{V} \mathbf{r}\times \partial_t\left(\rho\mathbf{u}\right) +
  \partial_t \mathbf{r}\times \rho\mathbf{u} \,{\rm d}V =
  \int_{V} \mathbf{r}\times \partial_t\left(\rho\mathbf{u}\right)\,{\rm
    d}V\, ,
\end{equation}
where we supposed that the volume V is held constant in the rotating frame.

In the context of star-planet systems, we are here primarily
interested in the rotational angular momentum of the star, and in the orbital
angular momentum of the planet. Both are defined through equation
\eqref{eq:amom_def}, respectively projected on the rotation axis and
on the normal to the orbital plane.

Combining equations \eqref{eq:jdot} and \eqref{eq:mom} we get
\begin{eqnarray}
    \dot{\boldsymbol{\mathcal{J}}} &=& \int_{V}  \mathbf{r} \times \left[\Div\left( -\rho\mathbf{u}\mathbf{u} +
    \mathbf{B}\mathbf{B}/4\pi - \mathbf{I}P_{t}\right)\right] \,
  {\rm d}V -2\int_V \rho  \mathbf{r} \times \left(\boldsymbol{\Omega}\times\mathbf{u}\right)\,
  {\rm d}V \nonumber \\  &+& \int_V \rho  \mathbf{r}\times\left(\boldsymbol{\Omega}\times \left(\boldsymbol{\Omega}\times\mathbf{r}\right)\right) \,
  {\rm d}V + \int_V \rho \mathbf{r}\times \mathbf{g}  \,
  {\rm d}V \, .
  \label{eq:jdot2}
\end{eqnarray}
In all the cases considered here, the gravity profile will be close to
symmetric in the volume of integration $V$, hence the last term of
Equation \ref{eq:jdot2} will be neglected. The centrifugal
contribution can be reworked through
\begin{equation}
  \label{eq:centrifugal}
  \int_V \rho \mathbf{r}\times \left(\boldsymbol{\Omega}\times \left(\boldsymbol{\Omega}\times\mathbf{r}\right)\right) \,
  {\rm d}V = \int_V \rho\boldsymbol{\Omega}\times \left(\mathbf{r}\times \left(\boldsymbol{\Omega}\times\mathbf{r}\right)\right) \,
  {\rm d}V \, . 
\end{equation}

In order to rewrite the first two terms of the right hand side of
\eqref{eq:jdot2}, we use the Levi-Civita permutation symbol $\varepsilon_{ijk}$ and
Einstein's summation notation
($(\mathbf{a}\times\mathbf{b})_{i}=\varepsilon_{ijk}a^{j}b^{k}$). If a tensor $\mathbf{C}$ is symmetric, we can write
\begin{equation*}
  \partial_{x_l}\left(\varepsilon_{ijk}x_jC_{kl}\right) =
  \varepsilon_{ijk}\left(C_{kl}\delta_{jl} +
    x_j\partial_{x_l}C_{kl}\right) =
  \varepsilon_{ijk}x_j\partial_{x_l}C_{kl}\, .
\end{equation*}
Since the three tensors in the first term of the right hand side of
\eqref{eq:jdot2} are symmetric, we can define ${\bf T} \equiv -\rho\mathbf{u}\mathbf{u} +
    \mathbf{B}\mathbf{B} - \mathbf{I}P_{t} $ and write
\begin{eqnarray}
  \left[\int_{V} \mathbf{r} \times \left[\Div \mathbf{T}\right]  \,
  {\rm d}V\right]_i &=&  \int_{V} \varepsilon_{ijk}x_j\partial_{x_l}T_{kl} \,
  {\rm d}V = \int_{V} \partial_{x_l}\left(\varepsilon_{ijk}x_jT_{kl}\right) \,
  {\rm d}V \nonumber \\ 
  &=& \int_{S} \varepsilon_{ijk}x_jT_{kl}n_{l} \,
  {\rm d}S \, ,
  \label{eq:first_term}
\end{eqnarray}
where $S$ is the surface bounding $V$ and $\mathbf{n}$ the normal to
this surface. Finally, we expand the Coriolis contribution by
conveniently introducing the following surface integral
\begin{eqnarray}
  A_i&\equiv& \int_S
              \left[\mathbf{r}\times \left(\boldsymbol{\Omega}\times\mathbf{r}\right)\right]_i
              \left(\rho\mathbf{u}\cdot{\rm d}\mathbf{S}\right) \nonumber
  \\ &=&  \int_S \varepsilon_{ijk} x_j \rho u_l\varepsilon_{kmn}\Omega_mx_nn_l{\rm
    d}S = \int_V \partial_{x_l}\left(\varepsilon_{ijk}\varepsilon_{kmn} \rho x_j
    u_l\Omega_mx_n\right) {\rm d}V \nonumber\\
  &=& \int_V  \rho u_l \partial_{x_l}\left(\left[\delta_{im}\delta_{jn} -
      \delta_{jm}\delta_{in}\right] 
    \Omega_m x_j x_n\right) {\rm d}V \nonumber\\
  &=& \int_V  \rho u_l \partial_{x_l}\left(\Omega_ix_nx_n -
      \Omega_jx_jx_i\right) {\rm d}V   \nonumber \\
  &=& \int_V  \rho \left( 2\Omega_ix_nu_n - \Omega_jx_ju_i - x_i\Omega_ju_j 
      \right)
      {\rm d}V  \, ,
 \label{eq:fourth_term}
\end{eqnarray}
where we used mass continuity equation $\partial_{x_l}\left(\rho u_l\right)=0$
that is satisfied in steady-state. We note that the Coriolis
contribution in equation \eqref{eq:jdot2} can be written
\begin{eqnarray}
  \mathcal{C}_i&\equiv&\left[-2\int_V \rho \mathbf{r}\times \left(\boldsymbol{\Omega}\times\mathbf{u}\right) \,
    {\rm d}V\right]_i = \int_V
-2\rho\varepsilon_{ijk}x_j\varepsilon_{kmn}\Omega_mu_n\,{\rm d}V
                   \nonumber \\
  &=&  \int_V -2\rho\left(\delta_{im}\delta_{jn} -
      \delta_{jm}\delta_{in}\right)\Omega_mu_nx_j\,{\rm d}V \nonumber \\
  &=& \int_V \rho\left(2\Omega_jx_ju_i-2\Omega_iu_jx_j \right)\,{\rm
      d}V \, .
  \label{eq:coriolis}
\end{eqnarray}
Combining equations \eqref{eq:fourth_term} and \eqref{eq:coriolis} we
obtain
\begin{equation}
  \label{eq:coriolis_final}
  \mathcal{C}_i = -A_i + \int_V \rho\left(u_i\Omega_jx_j -
    x_i\Omega_ju_j \right)\,{\rm d}V = -A_i + \left[\int_V
  \rho\boldsymbol{\Omega}\times\left(\mathbf{r}\times \mathbf{u}\right)\,{\rm
    d}V\right]_i\, .
\end{equation}
We can finally combine equations \eqref{eq:jdot2},
\eqref{eq:centrifugal} and \eqref{eq:coriolis_final} to obtain
\begin{equation}
  \label{eq:jdot_final}
  \dot{\mathcal{J}_i} = \int_S
  \varepsilon_{ijk}x_j\left[-\rho u_l\left(u_k+\varepsilon_{kmn}\Omega_mx_n\right)
    - P_t\delta_{kl} + B_lB_k/4\pi\right)n_l \,{\rm d}S + \left[\boldsymbol{\Omega}\times\int_V
  \rho \mathbf{r}\times\left(\mathbf{u}+\boldsymbol{\Omega}\times\mathbf{r}\right)\,{\rm
    d}V\right]_i \, .
\end{equation}
The surface integral corresponds to the flux of angular
momentum through the boundaries of the integration volume $V$, and the
second term appears to account for the rotating frame.

We consider here only the case where the rotation axis is normal to
the orbital plane. In this case the definitions of
the rotational and orbital
angular momentum coincide, albeit with a different integration volume
$V$. In both cases, the angular momentum component of interest is
$\mathcal{J}_z$, which is the component aligned with the rotation axis
$\boldsymbol{\Omega}/\Omega$. The second term of equation
\eqref{eq:jdot_final} vanishes for $\dot{\mathcal{J}}_z$, leaving only the
surface integral balancing the evolution of the angular momentum contained in volume
$V$. If the system is in steady state, or slowly evolving (which will
be justified here \textit{a posteriori}), $\dot{\mathcal{J}}_z\approx0$ and as a result
this surface integral is zero as well.

We now consider a volume $V$ bounded by two spherical surfaces $S_P$
and $S$, centered on the orbiting planet location, of spherical radii
$R_P$ and $R$ ($R_P$ being typically the planetary radius).
For a slowly evolving system (or a steady-state system),
using equation \eqref{eq:jdot_final}, we deduce that the 
torque applied to the planet can then be simply written \citep[see also][]{1970MNRAS.149..197M,Vidotto:2014kk}
\begin{equation}
  \label{eq:torq_planets}
  \mathcal{T} = \int_{S}
  \varepsilon_{ijk}x_j\left[\rho u_l\left(u_k+\varepsilon_{kmn}\Omega_mx_n\right)
    + P_t\delta_{kl} - B_lB_k/4\pi\right)n_l \,{\rm d}S\, .
\end{equation}
This final expression can then be rewritten in any desired system of
coordinates, for any surface $S$ enclosing the planet.

\end{document}